\documentclass[twocolumn,superscriptaddress,amsmath, amssymb, amsfonts,preprintnumbers,aps,prd,longbibliography,nofootinbib]{revtex4-1}

\renewcommand{\sec}[1]{\section{#1}}

\usepackage{graphicx}
\usepackage[dvipsnames]{xcolor}
\usepackage{dcolumn}
\usepackage{bm}
\usepackage{amstext}
\usepackage{amssymb}
\usepackage{subcaption}
\usepackage{amsmath}
\usepackage{caption}
\usepackage{ragged2e}
\usepackage{graphicx}
\usepackage{color}
\usepackage{bbold}
\usepackage{float}
\usepackage{delimset} 
\usepackage[colorlinks=true]{hyperref}
\usepackage{orcidlink}
\usetikzlibrary{decorations.pathmorphing,decorations.markings,positioning, shapes, snakes, arrows,backgrounds,patterns}
\usepackage{tikz-feynman}

\pgfdeclarelayer{foreground}
\pgfsetlayers{background,main,foreground}

\tikzset{ 
	graviton/.style={line width=.8pt, -latex,decorate, decoration={snake, segment length=4pt,amplitude=1.8pt, pre length=.1cm, post length=.25cm}},
	worldline/.style={gray, line width=1pt},
	worldlineBold/.style={black, line width=.6pt},
        background/.style={black,dotted,line width=1pt},
	zUndirected/.style={line width=1pt},
	zParticle/.style={line width=1pt,postaction={decorate},decoration={markings,mark=at position .6 with {\arrow[#1]{latex}}}},
	zParticleF/.style={line width=1pt,postaction={decorate}},
	cscalar/.style={line width=1pt,postaction={decorate},decoration={markings,mark=at position .6 with {\arrow[#1]{latex}}}},
	cscalar2/.style={line width=1pt,postaction={decorate},decoration={markings,mark=at position .8 with {\arrow[#1]{latex}}}},
	photon/.style={line width =.8pt, decorate, decoration={snake, segment length=3pt, amplitude=1.8pt,  pre length=.1cm, post length=.1cm}},
	 mid arrow/.style={postaction={decorate,decoration={
        markings,
        mark=at position .5 with {\arrow[#1]{latex}}}}} ,
        worlddot/.style={dotted, line width=.8pt},
	worlddot2/.style={dotted, line width=1pt},
  snake/.style={
      line width =.8pt,
      decorate,
      decoration={
        snake,
        segment length=5pt,
        amplitude=1.4pt,
        pre length=.0cm,
        post length=.0cm}
  },
  snakeDirected/.style={
    line width =.8pt,
    decorate,
    decoration={
      snake,
      segment length=5pt,
      amplitude=1.4pt,
      pre length=.0cm,
      post length=.1cm},
      postaction={decorate,decoration={markings,mark=at position 1 with {\arrow[xshift=4.5pt]{Latex[length=5.5pt,width=4pt]}}}}},
  shadow/.style={
    draw=white,
    line width=7pt,
    shorten <=7pt,
    shorten >=7pt
  },
  dottedline/.style={
    dotted, line width=1.15
  },
  zParticle/.style={line width=1pt,postaction={decorate},decoration={markings,mark=at position .5 with {\arrow[xshift=3.5pt]{Latex[length=5pt,width=4pt]}}}}
    }

\newcommand{\drawPerturbativeVertex}[1]{
  \begin{pgfonlayer}{foreground}
    \draw [fill=black, thick] (#1) circle (.045);
  \end{pgfonlayer}
}
\newcommand{\drawStaticPerturbativeVertex}[1]{
  \begin{pgfonlayer}{foreground}
    \draw [fill=Cerulean, thick,draw=Cerulean] (#1) circle (.045);
  \end{pgfonlayer}
}
\newcommand{\drawDottedLineLR}[3]{
  \draw[dottedline] ($(#1)-(#3,0)$) -- ($(#2)+(#3,0)$) ;
}

\newcommand{\drawGravitonLine}[2]{
    \draw [snake] (#1) -- (#2) ;
}
\newcommand{\drawStaticGravitonLine}[2]{
    \draw [snake,Cerulean] (#1) -- (#2) ;
}

\newcommand{\drawBHLineR}[2]{
    \draw [solid,thick] (#1) -- ($(#1)+(#2,0)$) ;
}
\newcommand{\drawStaticBHLineR}[2]{
    \draw [solid,thick,Cerulean] (#1) -- ($(#1)+(#2,0)$) ;
}

\newcommand{\drawStaticDirectedBHLineR}[2]{
    \draw [zParticle,Cerulean] (#1) -- ($(#1)+(#2,0)$) ;
}
\newcommand{\drawBHLine}[2]{
    \draw [solid,thick] (#1) -- (#2) ;
}
\newcommand{\drawStaticBHLine}[2]{
    \draw [solid,thick,Cerulean] (#1) -- (#2) ;
}
\newcommand{\drawCurvedBHLine}[4]{
    \draw [solid,thick] (#1) to[in=#4,out=#3,looseness=1] (#2) ;
}
\newcommand{\drawStaticCurvedBHLine}[4]{
    \draw [solid,thick,Cerulean] (#1) to[in=#4,out=#3,looseness=1.2] (#2) ;
}

\newcommand{\drawDottedLineR}[2]{
    \draw [dottedline] (#1) -- ($(#1)+(#2,0)$) ;
}
\newcommand{\drawDottedLine}[2]{
    \draw [dottedline] (#1) -- (#2) ;
}

\newcommand{\drawRadiativeGravitonLine}[2]{
    \draw [snake,WildStrawberry,line width=1.2pt] (#1) -- (#2) ;
}

\newcommand{\drawCurvedGravitonLine}[5][1.5]{
    \draw [snake] (#2) to[in=#5,out=#4,looseness=#1] (#3) ;
}
\newcommand{\drawStaticCurvedGravitonLine}[5][1.5]{
    \draw [snake,Cerulean] (#2) to[in=#5,out=#4,looseness=#1] (#3) ;
}

\newcommand{\drawRadiativeCurvedGravitonLine}[4]{
    \draw [snake,WildStrawberry,line width=1pt] (#1) to[in=#4,out=#3,looseness=1.5] (#2) ;
}

\DeclareFontFamily{OT1}{pzc}{}
\DeclareFontShape{OT1}{pzc}{m}{it}{<-> s * [1.350] pzcmi7t}{}
\DeclareMathAlphabet{\mathpzc}{OT1}{pzc}{m}{it}

\def\sym{n'}

\def\cO{\mathcal{O}}

\def\cF{\mathcal{F}}

\def\cI{\mathcal{I}}

\def\d{\mathrm{d}}
\def\eps{\epsilon}

\def\nn{\nonumber}

\newcommand{\mn}{{\mu\nu}}
\newcommand{\ab}{{\alpha\beta}}

\newcommand*{\vct}[1]{\boldsymbol{#1}}

\begin{document}

\preprint{HU-EP-26/18-RTG}

\title{A Runway to Dissipation of Angular Momentum via Worldline Quantum Field Theory}

\author{Gustav Uhre Jakobsen\,\orcidlink{0000-0001-9743-0442}} 
\email{gustav.uhre.jakobsen@physik.hu-berlin.de}
\affiliation{%
Institut f\"ur Physik und IRIS Adlershof, Humboldt-Universit\"at zu Berlin,
Zum Gro{\ss}en Windkanal 2, 12489 Berlin, Germany
}

\author{Kathrin Stoldt\, \orcidlink{0009-0001-4858-5218}}
\email{kathrin.stoldt@physik.hu-berlin.de}
\affiliation{%
Institut f\"ur Physik und IRIS Adlershof, Humboldt-Universit\"at zu Berlin,
Zum Gro{\ss}en Windkanal 2, 12489 Berlin, Germany
}

\begin{abstract}
We extend the worldline quantum field theory formalism to include a direct diagrammatic method of computing the total flux of angular momentum from a black hole scattering event in the post-Minkowskian regime.
Remarkably, except for subtle zero-frequency gravitons, the diagrammatic and integrational challenge is in a one-to-one correspondence with the analogous calculation of the black hole impulses --- and the well-developed WQFT methodologies for the impulse may thus be directly imported to this problem.
Zero-frequency gravitons appear in this calculation as a ``static'' integration region in addition to the ``dynamical'' region usually encountered for the impulse.
We show that a large class of static contributions can be organized systematically by introducing $n$-point functions referred to as ``static correlators". 
They reduce to a simple one-loop integral family which we compute explicitly using integration-by-parts relations and the method of differential equations.
In passing, our analysis shows that static contributions disappear in space-time dimensions $D>4$.
As a concrete application of our new method, we compute explicitly the $\cO(G^3)$ total flux of angular momentum reproducing known results.
Further, we apply the same method to electromagnetism where we compute the analogous $\cO(\alpha^3)$ result.
\end{abstract}
 
\maketitle 

\sec{Introduction}

The general relativistic two-body problem has recently regained a central role in theoretical physics due to its relevance to gravitational-wave physics, particularly in generating the waveforms regularly observed on Earth by the LIGO-Virgo-KAGRA observatories \cite{LIGOScientific:2016aoc,KAGRA:2021vkt,LIGOScientific:2025hdt,LIGOScientific:2025slb}. In addition, the prospect of promising future detectors enhances the need for high-precision theoretical predictions \cite{LISA:2017pwj,Punturo:2010zz,Ballmer:2022uxx}. 
This problem is fascinatingly simple in its idealized mathematical description:
For example, when the constituents are black holes (BHs), all dynamics is determined solely by Einstein's field equations and is therefore purely geometric.
On the other hand, the highly nonlinear nature of these equations makes them a significant computational challenge.
As a result, a wide range of computational frameworks, both analytical and numerical, have been developed to tackle this problem (most famously, perhaps, the effective-one-body approach~\cite{Buonanno:2000ef} and numerical relativity~\cite{Pretorius:2005gq,Scheel:2025jct} respectively).

In particular, the weak-field regime may be analyzed with analytic methods where in the non-relativistic post-Newtonian (PN) approximation one expands the field equations around Newton's potential and Kepler orbits for the BHs~\cite{Blanchet:2013haa,Porto:2026fsd,Brunello:2025gpf,Brunello:2026anu,Bini:2026dvn,Bini:2021gat,Bini:2022enm}.
Instead, in the relativistic post-Minkowskian (PM) expansion one expands around straight-line motion of the BHs and around flat Minkowski space-time for the metric~\cite{Damour:2016gwp,Damour:2017zjx,Bjerrum-Bohr:2022blt, Kosower:2022yvp}.
Generally, the PN approximation is ideal for bound motion while the PM one is suited for the unbound case.
Theoretically, all regimes of the two-body system are interesting (bound and unbound) though only the bound system is currently relevant for gravitational wave observations.
Local-in-time contributions to the two-body observables or potential may, however, straightforwardly be analytically continued between the two regimes~\cite{Kalin:2019rwq,Saketh:2021sri,Cho:2021arx}.

The weak-field description of the two-body system has greatly benefitted from methods originally developed for quantum field theory (QFT).
Firstly, in its description as an effective field theory where the intuitive notion of the compact objects as point-like particles is formalized~\cite{Goldberger:2004jt,Porto:2016pyg,Levi:2018nxp,Saketh:2022xjb}. 
At leading order, the compact objects are described by a minimal action and finite size effects appear as effective Wilson couplings perturbatively suppressed by the compactness of the object.
Spin and higher-order multipoles of the compact objects may also straightforwardly be incorporated into this framework~\cite{Levi:2015msa,Mandal:2023hqa,Mandal:2024iug}.
As finite-size effects are perturbatively suppressed all compact objects look the same at leading orders and we will simply refer to them as BHs in the rest of this work.

The PM scattering regime further benefits from QFT methodologies developed for high-precision computations of scattering amplitudes in high energy physics~\cite{Beneke:1997zp,Henn:2013pwa,Laporta:2000dsw}.
In particular, advanced integration techniques can be imported for ``classical loops'' which, essentially, at each order in perturbation theory captures the (non-trivial) classical limit of quantum loop integrals.
The worldline quantum field theory (WQFT) formalism~\cite{Mogull:2020sak,Jakobsen:2023oow,Haddad:2025cmw,Jakobsen:2022psy,Mogull:2025cfn,Gonzo:2026yha,Bohnenblust:2026ujk,Bastianelli:2021nbs,Wang:2022ntx,Comberiati:2022ldk,Bjerrum-Bohr:2025bqg,He:2025how,Bohnenblust:2023qmy,Bohnenblust:2025gir,Ben-Shahar:2025tiz,Ben-Shahar:2023djm} neatly captures this synergy between classical PM scattering and QFT and has been employed to compute scattering observables at unprecedented high orders in perturbation theory~\cite{Driesse:2024feo,Driesse:2024xad,Driesse:2026qiz} including, also, finite-size effects and spin~\cite{Jakobsen:2023ndj,Jakobsen:2023hig,Jakobsen:2023pvx,Haddad:2025cmw,Hoogeveen:2025tew}.
A closely related approach to WQFT is PM-EFT \cite{Kalin:2020mvi,Mougiakakos:2021ckm,Dlapa:2022lmu,Dlapa:2026oyq,Dlapa:2025biy} both of which use worldlines for the effective description of the BHs. 
In addition, a great amount of approaches using massive quantum fields to model the BHs exist~\cite{Bern:2025wyd,Bjerrum-Bohr:2018xdl,Cheung:2018wkq,Kosower:2018adc,Bern:2019nnu,Brandhuber:2023hhy,DiVecchia:2023frv,Aoude:2020onz,Damgaard:2021ipf,Damgaard:2023vnx,Herrmann:2021lqe,Herrmann:2021tct,Damgaard:2019lfh,Bjerrum-Bohr:2021vuf,Cheung:2020gyp}.

Impressively high orders in perturbation theory have been reached in calculating the impulse (5PM, i.e. four classical loops~\cite{Driesse:2024feo,Driesse:2024xad,Driesse:2026qiz,Bern:2025wyd,Dlapa:2026oyq,Dlapa:2025biy,Bini:2025vuk}) which describes both conservative effects encapsulated by the scattering angle and dissipative effects of radiation of energy and spatial momentum recoil.
Nevertheless, the impulse is not sufficient to describe one last important dissipative effect --- the loss of angular momentum. 
There, complete spinless results exist up to 3PM order~\cite{Damour:2020tta,Jakobsen:2021smu,Manohar:2022dea,DiVecchia:2022piu}, while partial results have been established at 4PM~\cite{Heissenberg:2025ocy}.
Further, high-order spin effects have been added at 2PM~\cite{Jakobsen:2021lvp,Alessio:2022kwv,Alessio:2025flu} and tidal and spin effects at 3PM~\cite{Jakobsen:2023hig,Jakobsen:2023pvx,Heissenberg:2023uvo,Heissenberg:2022tsn}.
One reason for radiation of angular momentum not having been computed to equally high orders as the impulse might be that the flux of angular momentum is sensitive to subtle effects linked to the gravitational memory effect~\cite{Damour:2020tta,Bini:2024rsy,Veneziano:2022zwh,DiVecchia:2022owy,Riva:2023xxm}.

In this paper we take first steps in preparing for the derivation of perturbative PM results for the (total) angular momentum flux at equally high perturbative orders as the impulse.
Within the WQFT formalism, we present a new simple formula from which the change in the impact parameter of each BH can be derived directly.
Using this ``impact parameter kick'' and the known impulse, it is a simple matter to compute the flux of angular momentum.
Apart from subtleties due to ``zero-frequency gravitons'', the computational challenge of the impact parameter kick intriguingly is no different from that of the impulse. 

Intuitively, zero-frequency gravitons~\cite{DiVecchia:2022owy,DiVecchia:2021ndb,Heissenberg:2021tzo,Riva:2023xxm,DeAngelis:2025vlf,Porto:2024cwd} appear in the impact parameter kick because of the ill-definedness of its time-component $\Delta t$ which naturally diverges.
These gravitons with frequencies $\omega_{\rm IR}\sim 1/\Delta t$ induce a coupling to the gravitational wave memory and necessitate an infrared regulator $\omega_{\rm IR}$.
This regulator which we take to define the ``static scale'' (in addition to the ``dynamical scale'' $q\sim1/b\gg\omega_{\rm IR}$, with $b$ being the impact parameter)
introduces new static integration regions.
We identify a large class of diagrams with a static contribution to the impact parameter kick related to, at least, linear memory effects. 
Their computational challenge may be captured by $n$-point static correlators which we show reduces to the computation of a specific one-loop integral which we solve explicitly.

The infinite (logarithmic) drift in the time component of the impact parameter kick implies that parts of this observable are badly behaved \cite{Caron-Huot:2023vxl, Bini:2024hme, Bini:2022wrq,Gralla:2021eoi,Gralla:2021qaf}.
Yet we observe that to the 3PM order considered in this work, this ill-definedness of the impact parameter is restricted to the conservative sector.
Thus, by focusing exclusively on dissipative effects, we avoid this intricacy.
This, however, presents no restriction as our desired observable, the change in angular momentum, is inherently dissipative and is thus only sensitive to the dissipative contributions of the impact parameter kick.

As a proof of concept and a first step in the quest of high-loop order results for the PM angular momentum flux, we compute explicitly the dissipative contributions to the impact parameter kick to third PM order.
This computation includes both static and dynamic contributions and reproduces well-known results in the literature.
As a further application, we apply the same techniques to electromagnetism (EM) and provide the EM total angular momentum flux to $\cO(\alpha^3)$ with the fine structure constant $\alpha$.
This latter analysis in EM complements other work using EM as a testing ground for GR~\cite{Bern:2023ccb,Bern:2021xze,Saketh:2021sri,Bini:2024pdp,Wang:2022ntx,Jakobsen:2023tvm}.

The structure of the paper is as follows.
First, in Sec.~\ref{sec:WEFTWQFT} we introduce the worldline effective field theory (WEFT) description of the weak-field limit of the general relativistic two-body problem and the WQFT approach to solving its equations of motion.
Second, in Sec.~\ref{sec:angularMomentum} we describe our method for efficiently extracting the angular momentum flux within the WQFT formalism.
Continuing, in Sec.~\ref{sec:ZFandStaticCorrelators} we analyze zero-frequency gravitons and static-region contributions to the angular momentum flux.
Then, in Secs.~\ref{sec:StaticAMFlux} and~\ref{sec:DynamicLoss} we discuss our 2PM and 3PM static computation and our 3PM dynamic computation respectively.
In the next two Secs.~\ref{sec:ResGR} and~\ref{sec:ResultEM} we present our $\cO(G^3)$ results for GR and $\cO(\alpha^3)$ results for EM respectively.
Finally, in Sec.~\ref{sec:conclusion}, we conclude with a summary and outlook for future work.

We use the mostly minus metric and put the speed of light to unity $c=1$.

\sec{WEFT of Compact Objects and the WQFT approach}
\label{sec:WEFTWQFT}
 
In this work, we follow the worldline effective field theory paradigm for describing BHs and other compact objects~\cite{Goldberger:2004jt,Porto:2016pyg,Levi:2018nxp,Saketh:2022xjb}.
More specifically, we employ the worldline quantum field theory approach which, in a simple manner, formalizes the application of quantum field theory methods to carrying out computations in this classical setting.
Working in the weak-field regime, we approximate the BHs (or neutron stars) by point-like particles.
This effective description necessitates a regularization of the UV physics of the BH and we employ dimensional regularization.

The minimal action for a point-like particle with a trajectory (worldline) $x_i^\mu(\tau)$ is its total proper time which in ``Brink - Di Vecchia - Howe'' form ~\cite{DESER1976369,Brink:1976sc,Mogull:2020sak} reads,
\begin{align}
    S_{{\rm BH}i}
    =
    -m_i
    \int \d\tau
    \big[
    \tfrac12 
    g_\mn(x_i(\tau))
    \dot x_i^\mu(\tau)
    \dot x_i^\nu(\tau)
    +\tfrac12
    \big]
    \ .
\end{align}
The final constant term $1/2$ may generally be dropped.
Spin degrees of freedom can straightforwardly be incorporated in this framework by the introduction of additional worldline fields (see e.g.~\cite{Haddad:2024ebn}).
Further, finite-size effects would be implemented as a tower of effective (Wilsonian) couplings perturbatively suppressed in the weak-field limit by the compactness of the BH.

The large-scale (compared with $Gm_i$) gravitational interaction of the BHs is described by the Einstein-Hilbert action,
\begin{align}
    S_{\rm EH}
    =
    -\frac{2}{\kappa^2}
    \int 
    \mathrm{d}^D x
    \sqrt{-g}
    R
    \, ,
\end{align}
with the Ricci curvature scalar $R$ and $\kappa^2=32 \pi G$.
The total action of our system then reads,
\begin{align}
    S
    =
    S_{\rm BH1}
    +
    S_{\rm BH2}
    +
    S_{\rm EH}
    +
    S_{\rm gf}
    \ .
\end{align}
The final term is a gauge-fixing and we have used both linear de Donder gauge and the optimized gauge of Ref.~\cite{Driesse:2024feo}.

In the spirit of WQFT, we expand the BH trajectories around their non-interacting, straight-line motion,
\begin{align}\label{eq:WLExp}
x_i^\mu(\tau)
=
b_i^\mu+v_i^\mu \tau
+
z_i^\mu(\tau)
\  ,
\end{align}
as well as the conventional flat-space background expansion of the metric,
\begin{align}
    g_\mn(x)
    =
    \eta_\mn
    +
    \kappa h_\mn(x)
    \ .
\end{align}
The perturbative, or \emph{dynamical}, fields are the worldline deflections $z^\mu_i(\tau)$ and the graviton $h_\mn(x)$.
The impact parameters $b_i^\mu$ and four velocities $v_i^\mu$ are instead background parameters.
We use proper time which implies that $v_i^2=1$ and we introduce the relative Lorentz factor of the two BHs $\gamma=v_1\cdot v_2$ as well as their relative velocity $v$ through the relation $(1-v^2)^{-1/2}=\gamma$.

The expansion of our action $S$ in the dynamical fields leads to the WQFT Feynman rules~\cite{Mogull:2020sak}.
The essential feature of WQFT, then, is that all observables of this classical two-body problem can be derived as the sum of one-point tree-level diagrams. 
Loop-diagrams represent quantum effects and are neglected.
Even so, within the WQFT tree diagrams, loop integrals nevertheless appear and the number of ``classical loops'' increases as the perturbative order --- in complete analogy with perturbative QFT.
To ensure causality in the classical observables, we apply the in-in formalism. As described in \cite{Jakobsen:2022psy,Kalin:2022hph},  in WQFT this amounts to the use of retarded propagators instead of the time symmetric Feynman prescription.

We assign a solid and a wiggly line to the BH deflections and the graviton respectively:
\begin{align}\label{WLprops}
    \begin{tikzpicture}[baseline={(current bounding box.center)}]
        \coordinate (in) at (-0.6,0);
        \coordinate (out) at (1.4,0);
        \coordinate (x) at (-.2,0);
        \coordinate (y) at (1.0,0);
        \draw [zParticle] (x) -- (y) node [midway, below] {$\omega$};
        \draw [dotted,thick] (in) -- (x);
        \draw [dotted,thick] (out) -- (y);
        \draw [fill] (x) circle (.06) node [above] {$\mu$} ;
        \draw [fill] (y) circle (.06) node [above] {$\nu$} ;
    \end{tikzpicture}
    &=\ 
    -i\frac{\eta^{\mu\nu}}{m_j}
    \frac{1}{(\omega+i0)^{2}}
    \ ,
    \\
    \begin{tikzpicture}[baseline={(current bounding box.center)}]
        \coordinate (in) at (-0.6,0);
        \coordinate (out) at (1.4,0);
        \coordinate (x) at (-.2,0);
        \coordinate (y) at (1.0,0);
        \draw [dotted,white] (x) -- (y) node [midway, below,black] {$k$};
        \drawGravitonLine{x}{y}
        \draw [dotted,white] (in) -- (x);
        \draw [dotted,white] (y) -- (out);
        \draw [fill] (x) circle (.06) node [above] {$\mu\nu$} ;
        \draw [fill] (y) circle (.06) node [above] {$\alpha\beta$} ;
    \end{tikzpicture}
    &=\ 
    i
    \frac{
        \eta_{\mu(\alpha}
        \eta_{\beta)\nu}
        -\tfrac1{D-2}
        \eta_\mn \eta_\ab
    }{(k^0+i0)^2-\vct{k}^2}
    \ .
\end{align}
Here, the WL frequency $\omega$ and graviton momentum $k^\mu$ are aligned with the causal flow which we generally take to flow from left to right (i.e. time runs from left to right).

The non-linearity of the Einstein Hilbert action introduces well-known $n$-graviton interaction vertices with any $n\ge3$.
Further, the graviton interacts with the BHs through worldline interaction vertices and, in the simple case of non-spinning BHs, they are given by a tower of interactions between a single graviton and any number (including zero) of WL deflections. We refer to \cite{Mogull:2020sak, Haddad:2024ebn, Haddad:2025cmw,Jakobsen:2023oow,He:2025how} for explicit vertex rules and relations obeyed by them.
The worldline vertices distinguish themselves in that they conserve only energy (the impact parameter $b_i^\mu$ breaks spatial translation invariance).
This leads to unconstrained spatial integrations on graviton momenta which leads to the classical loops.
The WQFT approach makes these loops manifest through the use of a dotted line which can be viewed to signify the inertial (straight-line, background) motion of the BHs.

\sec{Loss of Angular Momentum from WQFT}
\label{sec:angularMomentum}
In the WQFT framework no simple formula exists for the loss of angular momentum and instead it has been derived at LO via the waveform and the trajectory~\cite{Jakobsen:2021smu,Jakobsen:2021lvp,Mogull:2025cfn} and at NLO via linear response~\cite{Jakobsen:2023hig,Jakobsen:2023pvx}. 
However, it can be calculated as follows, once we have knowledge of the impact parameters $b_i^\mu$ and momenta $p_i^\mu=m_i v_i^\mu$ of the individual BHs:
\begin{align}
    L_i^\mn
    =
    2 b_i^{[\mu}
    p_i^{\nu]}
    \, .
\end{align}
We want to compute the dissipative changes in these orbital angular tensors $\Delta L_i^{\mu\nu}$ where we will generally use the notation that,
\begin{align}
    \Delta \cO = \cO_{\rm \infty}-\cO_{\rm -\infty}
    \ ,
\end{align}
with $\cO$ any background parameter and the subscript referring to evaluation at future infinity ($\infty$) or past infinity ($-\infty$).
The method that we will describe in a moment arrives at $\Delta L_i^\mu$ via knowledge of the change in its building blocks: $\Delta p_i^\mu$ and $\Delta b_i^\mu$.

First, let us construct the total angular momentum tensor $L^\mn$ and vector $L^\mu$ from the individual angular momentum tensors:
\begin{align}
    L^\mn
    =
    \sum_i
    L_i^\mn
    \ ,
    \qquad
    L^\mu
    =
    \frac1{2E}
    \eps^{\mu}_{\ \ab\nu}
    L^\ab P^\nu
    \ ,
\end{align}
Here $P^\mu=p_1^\mu+p_2^\mu$ is the total four momentum whose modulus $P^2=E^2$ is the total energy.
The (total) angular momentum vector is significant because it is invariant under translations of the coordinate system (which act as translations of $b_i^\mu$)~\cite{Jakobsen:2023oow,Jakobsen:2022zsx}.
This vector therefore represents a coordinate invariant measure of the total angular momentum which in the CoM frame where $\vct{P}=0$ coincides with the spatial components of $L^\mn$.
In the present work we will focus on translation invariant observables and we will therefore work with the vector $L^\mu$.

One finds the following expression for $L^\mu$ directly in terms of the impact parameters $b_i^\mu$ and the impulses $p_i^\mu$:
\begin{align}\label{eq:LDefinition}
    L^\mu
    =
    -\frac{1}{E}\eps^{\mu}_{\ \nu\ab}
     (b_2^\nu-b_1^\nu) p_1^\alpha p_2^\beta
    =
    L \hat L^\mu
    \ .
\end{align}
Here we also defined its (spatial) unit vector $\hat L^\mu$ with unit length $\hat L^2=-1$ and its modulus $L$.
Parts parallel to the velocities of $b_2^\mu-b_1^\mu$ do not contribute in this expression, suggesting that it depends only on the ``orthogonal impact parameter'',
\begin{align}\label{eq:OrthogonalImpactParameter}
    b_{\bot}^\mu
    =
    P_{\perp}^{\mu\nu}
    (b_{2\nu}-b_{1\nu})
    =
    b \hat b^\mu
    \ .
\end{align}
Here $P_{\perp}^\mn$ is a projector into the sub-space orthogonal to $v_i^\mu$,
\begin{align}
    P_{\perp}^\mn
    =
    \eta^\mn
    -
    \frac1{\gamma^2-1}
    \big(
        2\gamma v^{(\mu}_1 v^{\nu)}_2
        -
        v_1^\mu v_1^\nu
        -
        v_2^\mu v_2^\nu
    \big)
    \ ,
\end{align}
and, again, $\hat b^\mu$ is a spatial unit vector and $b$ is the modulus of $b_\perp^\mu$.
The orthogonal impact parameter $b_\perp^\mu$ is the physically relevant measure of the relative distance of the two BHs with its magnitude $b\sim1/q$ defining the relevant scale of the problem. Here $q$ is the modulus of the momentum transfer $q^\mu$, which is the dual variable to $b^\mu$ in Fourier space.
Finally the relation,
\begin{align}\label{eq:l}
    L= \frac{m_1 m_2 \sqrt{\gamma^2-1}b}{E}
    \ ,
\end{align}
connects the modulus of $b_\perp^\mu$ with that of $L^\mu$.

If one were to introduce intrinsic spins $S^\mn$ of the BHs, one would then focus on the total spin ${J^\mn=S^\mn+L^\mn}$. 
This inclusion is straightforward (see e.g.~\cite{Jakobsen:2022zsx}) and interesting but will be left for future work.
Since spinless scattering is confined to a plane, the angular momentum vector is restricted to pointing out of that plane.
We may thus focus exclusively on its modulus $L$. 
The main objective of this work is to compute $\Delta L$ which is purely dissipative and obeys $\Delta L^\mu=(\Delta L) \hat{L}^\mu$ .

Let us now turn to the ``building blocks" of $\Delta L$, namely $\Delta p_i^\mu$ and $\Delta b_i^\mu$.
As it is mainly the first of these, the impulse, which has been extensively studied \cite{Driesse:2024feo,Driesse:2024xad,Driesse:2026qiz,Bern:2025wyd}, the central quantity in this work shall be the change in the impact parameter, or, the ``impact parameter kick''.

As the main application of the impact parameter kick is in a derivation of the dissipative angular momentum flux, we may focus only on dissipative contributions.
As discussed in the introduction, this avoids consideration of its conservative components where, in particular, the components along the worldline velocities are ill-defined.
Furthermore, if the conservative change in the orthogonal impact parameter $b_\perp^\mu$ is desired, it is fully determined from the sole knowledge of the conservative impulse (and in turn the scattering angle).

How can we extract $\Delta b_i^\mu$ within the WQFT formalism? The conventional WQFT formula for the impulse reads:
\begin{align}\label{eq:impulse}
    \Delta p^\mu_i
    =
    -m_i 
    \lim_{\omega\to0}
    \omega^2 \langle z^\mu_i(\omega)\rangle
    \ .
\end{align}
This is the leading ``soft'' behavior in $\omega$ of the frequency space WL field.
As we will see, the subleading term in this expansion is related to $\Delta b^\mu_i$ and we write,
\begin{align}\label{eq:softExpansion}
	- \lim_{\omega\to0} \omega^2 \langle z_i^\mu(\omega) \rangle
	&=
	\frac{\Delta p_i^\mu}{m_i}
    +
	i\omega (
        \Delta b_{i,\rm dyn}^\mu
        +
        \omega^{-2\eps} 
        \Delta b_{i,\rm stat}^\mu
    )
    \nn
    \\
	&\hspace{4cm}
	+
	\dots\ ,
\end{align}
where the dots indicate terms of subleading order in $\omega$.
Intuitively, one may verify the identification of $\Delta b^\mu_i$ in this formula by an asymptotic Fourier transform from $\omega$ to $\tau$.
In this spirit, small $\omega$ behavior corresponds to asymptotically large ranges $\Delta t$ of the time component of the impact parameter, with $\omega \Delta t\sim1$.

A direct extraction of the impact parameter kick may be achieved by differentiation with respect to $\omega$ before the on-shell limit:
\begin{align}
    \Delta b_{i,{\rm dis}}^\mu
    =
    \lim_{\omega_{\rm IR}\to0}
    \frac{\d}{\d\omega_{\rm IR}}
    i\omega_{\rm IR}^2
    \langle z^\mu_i(\omega_{\rm IR})
    \rangle
    \bigg|_{\rm dis}
    \ .
    \label{eq:DeltaB}
\end{align}
Here, we have explicitly focused on dissipative effects and introduced a special notation for the external frequency $\omega_{\rm IR}$ which must be sent to zero.
As the $\omega_{\rm IR}\to0$ limit is non-trivial for $\Delta b^\mu_i$, we must consider integration regions which scale with $\omega_{\rm IR}$ in the course of the calculation. 
It is these regions which give rise to the split of Eq.~\eqref{eq:softExpansion} into a dynamical piece (dyn) and a static piece (stat).
In this respect, $\omega_{\rm IR}$ plays the role of an infrared scale.
As we are interested in asymptotic observables $\Delta t\gg b$ and, since $q\sim 1/b$, we have $\omega_{\rm IR}\ll q$.
Given that the wave memory describes a non-zero asymptotic offset of the metric, one can intuitively anticipate that $\omega_{\rm IR}$, corresponding to asymptotic times, couples to this effect. 

In concluding this section, let us note that $\Delta b_\perp^\mu$ has a simple, polynomial mass expansion in analogy to the known simple mass expansion of the impulse.
In fact, this may immediately be concluded from its direct relation to the WQFT one-point function which generally has such a simple, polynomial mass structure.
Concretely, one finds:
\begin{align}
    \Delta b_\perp^\mu
    =
    \sum_{n=1}^{\infty}
    \sum_{k=0}^n \frac{G^n m_1^{k} m_2^{n-k}}{b^{n-1}}
    c_{(n,k)}^\mu
    \ ,
\end{align}
where $c_{n,k}^\mu$ are dimensionless vector coefficients which apart from their directional dependence are functions only of $\gamma$.
Note that dissipative effects appear only for $n\ge2$ in this equation.
From Eq.~\eqref{eq:LDefinition}, one sees that,
\begin{align}\label{eq:EL}
    \Delta (EL^\mu)
    =
    \Delta (E L)\hat L^\mu
    =
    -\epsilon^\mu{}_{\nu\ab}
    \Delta(b_\perp^\nu p_1^\alpha p_2^\beta)
    \ .
\end{align}
This equation implies that $\Delta(E L)$ will have a polynomial mass expansion which, instead, is not the case individually for its building blocks $\Delta E$ and $\Delta L$.
It is therefore convenient to present results for the loss of angular momentum through the intermediate observable $\Delta(EL)$.

\sec{Zero Frequency Gravitons and Static Correlators}\label{sec:ZFandStaticCorrelators}

Above we have seen that, to compute $\Delta b^\mu_i$ or $\Delta p^\mu_i$, we must take the limit $\omega_{\rm IR}\to0$ of the worldline one-point function.
For the impulse this limit has (as of yet) not exhibited any non-trivial regions.
This, however, is different for the impact parameter kick and, equivalently, for the dissipation of angular momentum.
For non-zero, but small, frequency, we have two scales $q\sim1/b$ and $\omega_{\rm IR}\ll q$.

In the desired $\omega_{\rm IR}\to0$ limit in eq.~\eqref{eq:DeltaB}, we find non-trivial integration regions where loop momenta scale with $\omega_{\rm IR}$.
Such ``zero-frequency gravitons'' are already well-known from previous work ~\cite{DiVecchia:2022owy}.
Generally, we find two classes of loop momenta which we refer to as dynamical or static:
\begin{align}\label{eq:scalings}
    \begin{aligned}
    \ell_{\rm dyn}^\mu
    &=
    (\ell^0_{\rm dyn},\vct{\ell}_{\rm dyn})
    \sim (q,q)
    \ ,
    \\
    \ell_{\rm stat}^\mu
    &=
    (\ell^0_{\rm stat},\vct{\ell}_{\rm stat})
    \sim (\omega_{\rm IR},\omega_{\rm IR})
    \ .
    \end{aligned}
\end{align}
The dynamical gravitons scale as the external scale $q$ and are the only kinds of gravitons considered in calculations of the impulse.
Instead, the static gravitons scale with the infrared scale $\omega_{\rm IR}$ which is vanishingly small.
We apply the same nomenclature to frequencies of the BHs as they may be thought of as one-dimensional momenta with a time component only.

We do not consider mixed scalings of the gravitons in this work (i.e. something like $\ell^\mu\sim(\omega_{\rm IR},q)$) because we focus only on dissipative effects and a graviton with such a mixed scaling cannot go on-shell.
This may change at higher perturbative orders where a ``conservative'' graviton with mixed scalings could combine with a different on-shell graviton and give rise to dissipative effects.

Let us consider under which conditions a WQFT one-point function $\langle z^\mu(\omega_{\rm IR})\rangle$ can have a non-vanishing zero-frequency contribution (in the $\omega_{\rm IR}\to0$ limit) and how it scales with $\omega_{\rm IR}$.
Take as an example the following diagram, 
\begin{align}\label{eq:exampleDiagram}
    \begin{tikzpicture}[baseline={(0,.4)}]
        \drawStaticGravitonLine{0,0}{0,.9}
        \drawGravitonLine{.6,0}{.6,.9}
        \drawStaticGravitonLine{1.2,0}{1.2,.9}
        \drawDottedLineR{0,0}{-.4}
        \drawDottedLineR{.6,0}{1}
        \drawDottedLineR{.6,.9}{-1}
        \drawBHLineR{0,0}{.6}
        \drawStaticBHLineR{.6,.9}{1}
        \drawPerturbativeVertex{0,0}
        \drawStaticPerturbativeVertex{0,.9}
        \drawPerturbativeVertex{.6,0}
        \drawPerturbativeVertex{.6,.9}
        \drawStaticPerturbativeVertex{1.2,0}
        \drawStaticPerturbativeVertex{1.2,.9}
        \node[right] at (1.6,.9) {$z^\mu(\omega_{\rm IR})$} ;
    \end{tikzpicture}
    \ ,
\end{align}
where we highlighted certain graviton momenta and WL frequencies in blue which we assume to be static (in this example).
Further, in a vertex rule, if it is connected to a dynamic momentum (or frequency), this dynamic variable will dominate over any vanishingly small static variables and we draw the vertex in black.
Instead, if a vertex is connected only to static propagators, it will be a function only of static variables and we draw it in blue.

With these simple assumptions, the example diagram~\eqref{eq:exampleDiagram} will factorize into three parts: Namely, each connected static subdiagram as well as each connected dynamic subdiagram will factorize from each other and form separate integrals.
This happens as --- after assuming the scalings of Eq.~\eqref{eq:scalings} --- one should Taylor expand the integrand in $\omega_{\rm IR}/q\ll1$ and in this process all static momenta must factorize from the dynamic momenta.
In order for a static subdiagram to be non-zero, it must depend on the static scale $\omega_{\rm IR}$ which is only the case if it is connected with the external line.
Thus, in particular, the left-most subdiagram of eq.~\eqref{eq:exampleDiagram} is scaleless and must vanish.

A non-vanishing static contribution must then come only from a diagram with a single static sub-diagram which connects the outgoing line with some number of incoming dynamic subdiagrams.
This situation is drawn schematically as,
\begin{align}\label{eq:GeneralDiagram}
    \begin{tikzpicture}[baseline={(0,0)}]
    \coordinate (top) at (0,.64cm) ;
    \coordinate (topOut) at (.7,.64cm) ;
    \coordinate (bot) at (0,-.64cm) ;
    %
    %
    \coordinate (topA)at (-.7,.64cm) ;
    \coordinate (topB)at (-.7,1.1cm) ;
    \coordinate (topAA)at(-1.,.64cm) ;
    \coordinate (topBB)at(-1.,1.1cm) ;
    %
    %
    \coordinate (botA)at (-.7,-.64cm) ;
    \coordinate (botB)at (-.7,-1.1cm) ;
    \coordinate (botAA)at(-1.,-.64cm) ;
    \coordinate (botBB)at(-1.,-1.1cm) ;
    %
    %
    \coordinate (topCC)at(-1.,.2cm) ;
    \coordinate (botCC)at(-1.,-.2cm) ;
    \coordinate (topC)at (-.7,.2cm) ;
    \coordinate (botC)at (-.7,-.2cm) ;
    %
    %
    \drawStaticDirectedBHLineR{top}{.7} 
    \drawDottedLineR{bot}{.7} 
    %
    %
    \drawStaticCurvedGravitonLine{0,0}{topCC}{160}{0}
    \drawStaticCurvedGravitonLine{0,0}{botCC}{200}{0}
    %
    %
    \drawStaticCurvedBHLine{top}{topA}{180}{0}
    \drawStaticBHLine{topA}{topAA}
    \drawStaticCurvedBHLine{top}{topB}{160}{0}
    \drawStaticBHLine{topB}{topBB}
    \drawStaticCurvedBHLine{bot}{botA}{180}{0}
    \drawStaticBHLine{botA}{botAA}
    \drawStaticCurvedBHLine{bot}{botB}{220}{0}
    \drawStaticBHLine{botB}{botBB}
    %
    %
    \draw[line width=.8pt,fill=white] (topAA) circle (.17cm) ;
    \draw[pattern=north east lines] (topAA) circle (.17cm) ;
    \draw[line width=.8pt,fill=white] (topBB) circle (.17cm) ;
    \draw[pattern=north east lines] (topBB) circle (.17cm) ;
    \draw[line width=.8pt,fill=white] (botAA) circle (.17cm) ;
    \draw[pattern=north east lines] (botAA) circle (.17cm) ;
    \draw[line width=.8pt,fill=white] (botBB) circle (.17cm) ;
    \draw[pattern=north east lines] (botBB) circle (.17cm) ;
    \draw[line width=.8pt,fill=white] (topCC) circle (.17cm) ;
    \draw[pattern=north east lines] (topCC) circle (.17cm) ;
    \draw[line width=.8pt,fill=white] (botCC) circle (.17cm) ;
    \draw[pattern=north east lines] (botCC) circle (.17cm) ;
    %
    %
    \node[rotate=90] at ($(topA)!0.5!(topB)$) {\scriptsize ...} ;
    \node[rotate=90] at ($(botA)!0.5!(botB)$) {\scriptsize ...} ;
    \node[rotate=90] at ($(topC)!0.5!(botC)$) {\scriptsize ...} ;
    %
    %
    \draw[line width=.8pt,fill=white,draw=Cerulean] (0,0) ellipse (.26cm and .67cm);
    \draw[pattern=north east lines,pattern color=Cerulean,draw=Cerulean] (0,0) ellipse (.26cm and .67cm);
    %
    %
    \draw[decorate,decoration={brace},thick] ($(topAA)+(-.3,-.15)$) -- ($(topBB)+(-.3,.15)$) node[midway, left=3pt] {$n$} ;
    \draw[decorate,decoration={brace},thick] ($(botBB)+(-.3,-.15)$) -- ($(botAA)+(-.3,.15)$) node[midway, left=3pt] {$\sym$} ;
    \draw[decorate,decoration={brace},thick] ($(botCC)+(-.3,-.15)$) -- ($(topCC)+(-.3,.15)$) node[midway, left=3pt] {$l$} ;
    \node[right] at (topOut) {$z^\mu(\omega_{\rm IR})$} ;
    \end{tikzpicture}
\end{align}
Note, again, that we generally draw diagrams with causality (i.e. time) flowing from left to right.
In this diagram, the (connected) static subdiagram has $(n+l+\sym)$ incoming external legs connected with $n$ dynamic $z_1^\mu$ subdiagrams, $l$ dynamic $h_\mn$ subdiagrams and $\sym$ dynamic $z_2^\mu$ subdiagrams.
Since we are studying classical effects, all subdiagrams (symbolized by shaded blobs) must have tree topologies and must, by assumption, be connected.
Apart from that, they can generally have a non-trivial structure and contributions at any order in perturbation theory.

As discussed, the general diagram in Eq.~\eqref{eq:GeneralDiagram} factorizes into the connected $(n+l+\sym+1)$-point function --- which we will refer to as a ``static correlator'' --- which is contracted with $(n+l+\sym)$ dynamic one-point functions.
The dynamic one-point insertions are independent of $\omega_{\rm IR}$ which will only appear in the static correlator (at leading order in $\omega_{\rm IR}\to0$).

Let us analyze the generic static correlator which we identify as
\begin{align}\label{eq:staticCorrelator}
    \int_{\{\omega_i,\omega'_i,\ell_i\}}
\begin{tikzpicture}[baseline={(0,0)}]
    \coordinate (top) at (0,.64cm) ;
    \coordinate (topOut) at (.7,.64cm) ;
    \coordinate (bot) at (0,-.64cm) ;
    %
    %
    \coordinate (topA)at (-.7,.64cm) ;
    \coordinate (topB)at (-.7,1.1cm) ;
    \coordinate (topAA)at(-.85,.64cm) ;
    \coordinate (topBB)at(-.85,1.1cm) ;
    %
    %
    \coordinate (botA)at (-.7,-.64cm) ;
    \coordinate (botB)at (-.7,-1.1cm) ;
    \coordinate (botAA)at(-.85,-.64cm) ;
    \coordinate (botBB)at(-.85,-1.1cm) ;
    %
    %
    \coordinate (topCC)at(-.85,.2cm) ;
    \coordinate (botCC)at(-.85,-.2cm) ;
    \coordinate (topC)at (-.7,.2cm) ;
    \coordinate (botC)at (-.7,-.2cm) ;
    \node[right] at (topOut) {$\omega_{\rm IR}$} ;
    %
    %
    \drawStaticDirectedBHLineR{top}{.7} 
    \drawDottedLineR{bot}{.7} 
    %
    %
    \drawStaticCurvedGravitonLine{0,0}{topCC}{160}{0}
    \drawStaticCurvedGravitonLine{0,0}{botCC}{200}{0}
    %
    %
    \drawStaticCurvedBHLine{top}{topA}{180}{0}
    \drawStaticBHLine{topA}{topAA}
    \drawStaticCurvedBHLine{top}{topB}{160}{0}
    \drawStaticBHLine{topB}{topBB}
    \drawStaticCurvedBHLine{bot}{botA}{180}{0}
    \drawStaticBHLine{botA}{botAA}
    \drawStaticCurvedBHLine{bot}{botB}{220}{0}
    \drawStaticBHLine{botB}{botBB}
    %
    %
    \node[rotate=90] at ($(topA)!0.5!(topB)$) {\scriptsize ...} ;
    \node[rotate=90] at ($(botA)!0.5!(botB)$) {\scriptsize ...} ;
    \node[rotate=90] at ($(topC)!0.5!(botC)$) {\scriptsize ...} ;
    %
    %
    \draw[line width=.8pt,fill=white,draw=Cerulean] (0,0) ellipse (.26cm and .67cm);
    \draw[pattern=north east lines,pattern color=Cerulean,draw=Cerulean] (0,0) ellipse (.26cm and .67cm);
    %
    %
    \node[left] at (topBB) {$\omega_1$} ;
    \node[left] at (topAA) {$\omega_n$} ;
    \node[left] at (botAA) {$\omega_1'$} ;
    \node[left] at (botBB) {$\omega_{\sym}'$} ;
    \node[left] at (topCC) {$\ell_1^\mu$} ;
    \node[left] at (botCC) {$\ell_l^\mu$} ;
\end{tikzpicture}\, ,
\end{align}
where we make use of the notation,
\begin{align}
\int_\ell:=\int \frac{\mathrm{d}^D\ell}{(2\pi)^D}\ ,\qquad \int_\omega:=\int\frac{\mathrm{d}\omega}{(2\pi)}\ .
\end{align}
This definition extends to each frequency and momentum in the set $\{\omega_i,\omega'_i,\ell_i\}$.
To avoid an overload of notation in equation (\ref{eq:staticCorrelator}), we have omitted external indices of the fields but include their frequencies and momenta.
We do not amputate external propagators and we integrate over all incoming ones.
This integration is simply the usual integration on all internal momenta and frequencies which has now been exposed by the factorization of the diagram in eq.~\eqref{eq:GeneralDiagram}.
By assumption (and diagrammatically signified by the blue coloring), all frequencies and momenta are static and thus scale with $\omega_{\rm IR}$. 
After the diagram has been expanded in this limit one may however integrate over their full ranges again (as dictated by the method of regions~\cite{Smirnov:2001in,Smirnov:1999bza,Beneke:1997zp}).

The static correlator depicted in eq.~\eqref{eq:staticCorrelator} does, by assumption, not depend on the dynamical scale $q$.
Its dependence on its only scale $\omega_{\rm IR}$ can thus be determined from dimensional analysis.
The dimension of a general static correlator may easily be determined and, taking into account integrations on incoming momenta or frequencies, for the generic static correlator~\eqref{eq:staticCorrelator} we find:
\begin{align}
    m^{-n-\sym}\kappa^{l} \omega_{\rm IR}^{2l-2-2l\eps} 
    \ .
\end{align}
Here $m$ is a generic mass (could be any of $m_i$).
While this completely determines the dimension of the correlators, it does not yet determine their dependence on $\omega_{\rm IR}$ since the frequency combines with $G$ and $m$ into the dimensionless variable $G m \omega_{\rm IR}^{1-2\eps}$.
It is this variable which controls the perturbative expansion of the correlators.

For zero incoming dynamical gravitons (i.e. $l=0$) the diagram must have at least one internal graviton.
This is because no interaction vertex exists without incoming gravitons and thus we must have at least one classical loop.
Therefore, in this case the static correlator must scale with at least one overall $G$ and we find:
\begin{align}\label{eq:countingDim}
    m^{-n-\sym}\omega_{\rm IR}^{-2}
    \big(
        (Gm\omega_{\rm IR}^{1-2\eps}) +  (Gm\omega_{\rm IR}^{1-2\eps})^2+\dots
    \big)
    \ .
\end{align}
Thus, the leading order in $G$ of such a diagram scales as $\omega_{\rm IR}^{-1-2\epsilon}$ with all other contributions suppressed in the limit $\omega_{\rm IR}\to0$.
Further $\omega_{\rm IR}^{-1-2\epsilon}$ is exactly the right behavior for contributions to $\Delta b_i^\mu$ (see eq.~\eqref{eq:softExpansion}) and this class of static correlators therefore has non-zero contributions to the impact parameter kick.

Further, if there are any number $l>0$ of gravitons, we find
\begin{align}
    m^{-n-\sym} \kappa^l \omega_{\rm IR}^{2(l-1-l\eps)}
    (1+Gm \omega_{\rm IR}^{1-2\eps} +\dots)\ ,
\end{align}
which always scales in a subleading way compared to $\omega_{\rm IR}^{-1}$ because $2(l-1)\ge0$ and therefore does not contribute to the impact parameter kick.
We note that in this case there may be contributions without classical loops which is the reason for including the first $G$-independent term in the brackets (however only for $l=1$ in the current spinless and structureless context).

This dimensional analysis shows that only a few static correlators are relevant: They must have only external WLs and only one single internal graviton.
Following this observation we encounter two types of static $(n+\sym+1)$-point correlators:
\begin{subequations}\label{eq:staticCors}
\begin{align}\label{generalstaticSnakesMushrooms}
    &D_{12}^{\mu\rho_1\dots\rho_n;\rho'_1\dots\rho'_{\sym}}
    =
    \omega_{\rm IR}^{1-2\epsilon}
    \!\!\int_{\{\omega_i,\omega'_i\}}\!\!\!
\begin{tikzpicture}[baseline={(0,.2)}]
    \coordinate (top) at (0,.64cm) ;
    \coordinate (bot) at (0,-.64cm) ;
    \coordinate (topR)at (.7,.64cm) ;
    %
    %
    \coordinate (topA)at (-.7,.64cm) ;
    \coordinate (topB)at (-.7,1.1cm) ;
    \coordinate (topAA)at(-.85,.64cm) ;
    \coordinate (topBB)at(-.85,1.1cm) ;
    %
    %
    \coordinate (botA)at (-.7,-.64cm) ;
    \coordinate (botB)at (-.7,-.18cm) ;
    \coordinate (botAA)at(-.85,-.64cm) ;
    \coordinate (botBB)at(-.85,-.18cm) ;
    %
    %
    \drawStaticGravitonLine{bot}{top}
    %
    %
    \drawStaticDirectedBHLineR{top}{.7} 
    \drawDottedLineR{bot}{.7} 
    %
    %
    \drawStaticCurvedBHLine{top}{topA}{180}{0}
    \drawStaticBHLine{topA}{topAA}
    \drawStaticCurvedBHLine{top}{topB}{160}{0}
    \drawStaticBHLine{topB}{topBB}
    \drawStaticCurvedBHLine{bot}{botA}{180}{0}
    \drawStaticBHLine{botA}{botAA}
    \drawStaticCurvedBHLine{bot}{botB}{160}{0}
    \drawStaticBHLine{botB}{botBB}
    %
    %
    \node[rotate=90] at ($(topA)!0.5!(topB)$) {\scriptsize ...} ;
    \node[rotate=90] at ($(botA)!0.5!(botB)$) {\scriptsize ...} ;
    %
    %
    %
    %
    \node[left] at (topBB) {$z^{\rho_1}(\omega_1)$} ;
    \node[left] at (topAA) {$z^{\rho_n}(\omega_n)$} ;
    \node[left] at (botBB) {$z^{\rho'_1}(\omega_1')$} ;
    \node[left] at (botAA) {$z^{\rho'_{\sym}}(\omega_{\sym}')$} ;
    \node[below] at (topR) {$z^\mu(\omega_{\rm IR})$} ;
    %
    %
    \drawStaticPerturbativeVertex{bot}
    \drawStaticPerturbativeVertex{top}
\end{tikzpicture},
\\
    &D_{11}^{\mu\rho_1\dots\rho_n;\rho'_1\dots\rho'_{\sym}}
    =
    \omega_{\rm IR}^{1-2\epsilon}
    \!\!\int_{\{\omega_i,\omega'_i\}}
    \nn\\
    &\hspace{2cm}\times
\begin{tikzpicture}[baseline={(0,.6)}]
    \coordinate (top) at (0,.64cm) ;
    \coordinate (bot) at (-2.2,.64cm) ;
    \coordinate (topR)at (.7,.64cm) ;
    %
    %
    \coordinate (topA)at (-.7,.94cm) ;
    \coordinate (topB)at (-.7,1.4cm) ;
    \coordinate (topAA)at(-.85,.94cm) ;
    \coordinate (topBB)at(-.85,1.4cm) ;
    %
    %
    \coordinate (botA)at (-2.9,.64cm) ;
    \coordinate (botB)at (-2.9,1.1cm) ;
    \coordinate (botAA)at(-3.05,.64cm) ;
    \coordinate (botBB)at(-3.05,1.1cm) ;
    %
    %
    \drawStaticCurvedGravitonLine[1]{bot}{top}{-90}{-90}
    %
    %
    \drawStaticDirectedBHLineR{top}{.7} 
    \drawDottedLine{bot}{top} 
    %
    %
    \drawStaticCurvedBHLine{top}{topA}{180}{0}
    \drawStaticBHLine{topA}{topAA}
    \drawStaticCurvedBHLine{top}{topB}{160}{0}
    \drawStaticBHLine{topB}{topBB}
    \drawStaticCurvedBHLine{bot}{botA}{180}{0}
    \drawStaticBHLine{botA}{botAA}
    \drawStaticCurvedBHLine{bot}{botB}{160}{0}
    \drawStaticBHLine{botB}{botBB}
    %
    %
    \node[rotate=90] at ($(topA)!0.5!(topB)$) {\scriptsize ...} ;
    \node[rotate=90] at ($(botA)!0.5!(botB)$) {\scriptsize ...} ;
    %
    %
    %
    %
    \node[left] at (topBB) {$z^{\rho_1}(\omega_1)$} ;
    \node[left] at (topAA) {$z^{\rho_n}(\omega_n)$} ;
    \node[left] at (botBB) {$z^{\rho'_1}(\omega_1')$} ;
    \node[left] at (botAA) {$z^{\rho'_{\sym}}(\omega_{\sym}')$} ;
    \node[below] at (topR) {$z^\mu(\omega_{\rm IR})$} ;
    %
    %
    \drawStaticPerturbativeVertex{bot}
    \drawStaticPerturbativeVertex{top}
\end{tikzpicture} .
\end{align}
\end{subequations}
In the definition of $D_{1i}^{\cdots}$ we scale out the known behavior of $\omega_{\rm IR}$.
Furthermore, to have dissipative effects, $m$ must be non-zero.
Interestingly, it may be verified that in the low-velocity limit $v\to0$, where $v_2^\mu\to v_1^\mu$ the dissipative part of each type of correlator is identical (and in this way $D_{11}^{\cdots}$ may be derived from $D_{12}^{\cdots}$).
Further we note that the correlators $D_{11}^{\cdots}$ were already studied in EM in Ref.~\cite{Jakobsen:2023tvm} where they described local self-interaction. 

The general integral family for the static correlators reads:
\begin{align}
    \begin{aligned}
   &\mathcal{I}_{\{n_i,n_i^\prime\}}= \\
   &\int_{\{\omega_i,\omega_i',\ell\}}
    \frac{
        \delta(\omega_{\rm IR}-\ell\cdot v_1-{\textstyle\sum_i}\omega_i)
    }{
        \prod_{i=1}^n (\omega_i+i0)^{n_i}
    }
    \frac{
    \delta(\ell\cdot v_a-{\textstyle\sum_i}\omega_i')
    }{
        \prod_{i=1}^{\sym} (\omega_i'+i0)^{n_i'}
    }
    \frac{1}{\ell^2}\, .
    \end{aligned}
\end{align}
Note that the exponential Fourier factor which would have introduced the dynamic scale has disappeared due to the static scaling. 
The integrals on $\omega_i$ and $\omega'_i$ resemble one-dimensional $(n-1)$ or $(\sym-1)$-dimensional loop integrals respectively.
They have only one scale ($(\omega_{\rm IR}-\ell\cdot v_1)$ and $\ell\cdot v_a$ respectively) and their dependence on that scale can thus be directly determined from dimensional analysis.
Apart from that, their evaluation is simply a matter of contour integration.
Analyzing the diagram, one expects only powers $n_i,n_i'\in\{1,2\}$ which ensures convergence.

After evaluation of the frequency integrations (and allowing for the more general case of tensor integrals) we arrive at the family,
\begin{align}\label{eq:statInts}
    &
    \mathcal{I}^{\mu_1\dots\mu_k}_{(\nu_1\nu_2\nu_3)}(\gamma,\epsilon,\omega_{\rm IR})
    =
    \\
    &\qquad\qquad
    \int_\ell
    \frac{
        \ell^{\mu_1}
        \dots
        \ell^{\mu_k}
        }{
        \ell^{2\nu_1}
        (\ell\cdot v_2+i0)^{\nu_2}
        (\omega_{\rm IR}-\ell\cdot v_1+i0)^{\nu_3}
    }
    \, ,
    \nn
\end{align}
with tensor indices $\mu_i$ and propagator power indices $\nu_i$. 
We note a clear similarity of this integral family to the infrared one-loop integrals considered in Ref.~\cite{Heissenberg:2021tzo}.
The tensor integrals appear, here, because we are computing $(n+\sym+1)$-point correlation functions and the number of tensors $k$ will at most be the number of points $(n+\sym+1)$.

Let us analyze the pole structure of the static integrals in Eq.~\eqref{eq:statInts}. 
First we notice that only the pole of the third propagator is captured when closing the contour in the upper half plane.
If one adds to this integral the same integral with the opposite $i\epsilon$-prescription on the third propagator, the new addition obviously vanishes. 
However, it allows one to rewrite the integral so that the third propagator is cut and also implies $\nu_3>0$.
Further the integration on components of $\ell^\mu$ orthogonal to $v_i^\mu$ requires $\nu_1>0$ because otherwise the integral will be scaleless.

With tensor reduction, the integrals are reduced to scalars $\mathcal{I}_{\nu_1\nu_2\nu_3}$ and with IBP reduction they are further reduced to two master integrals given by,
\begin{align}   
    \vec{\mathcal{I}}(\gamma,\epsilon)
    =
    \omega_{\rm IR}^{2\epsilon}
    \begin{pmatrix}
        \ 
        \mathcal{I}_{(111)}(\gamma,\epsilon,\omega_{\rm IR})
        \hphantom{/\omega_{\rm IR}}
        \\[4pt]
        \ 
        \mathcal{I}_{(101)}(\gamma,\epsilon,\omega_{\rm IR})/\omega_{\rm IR} 
    \end{pmatrix}
    \ .
\end{align}
The factors of $\omega_{\rm IR}$ are inserted such that the vector $\vec{I}(\gamma,\epsilon)$ is independent of $\omega_{\rm IR}$.
The integration problem of the static correlators has thus been reduced to the computation of two master integrals. 

We use the approach of differential equations to solve the static master integrals,
\begin{align}
\frac{\mathrm{d}}{\mathrm{d}\gamma}
\vec{\cI}(\gamma,\epsilon)
=
M(\gamma,\epsilon)
\vec{\cI}(\gamma,\epsilon)
\ ,
\end{align}
with the simple matrix:
\begin{align}
M(\gamma,\epsilon)
=
-\frac{1-2\epsilon}{\gamma^2-1}
 \begin{pmatrix}
 1&\gamma
 \\
 0&0
\end{pmatrix}\ .
\end{align}
This matrix can be put in canonical form, which allows us to factor out the dependence on $\epsilon$ and to solve the differential equation order by order in the dimensional regulator.
Thereafter we transform back to the basis of $\vec{\cI}(\gamma,\epsilon)$ which are now determined up to boundary constants.

We follow the common choice in PM literature of fixing the boundary constants in the low-velocity PN limit where $v\to0$. 
In this limit, we find two kinds of static, zero-frequency gravitons which are analogous to the kind of regions usually found for dynamical gravitons:
Namely potential and radiative static gravitons with scalings as,
 \begin{align}
    \begin{aligned}
    \ell_{\rm pot.stat.}^\mu
    &=
    (\ell^0_{\rm pot.stat.},\vct{\ell}_{\rm pot.stat.})
    \sim 
    \frac{\omega_{\rm IR}}{v}
    (v,1)
    \ ,
    \\
    \ell_{\rm rad.stat.}^\mu
    &=
    (\ell^0_{\rm rad.stat.},\vct{\ell}_{\rm rad.stat.})
    \sim 
    \frac{\omega_{\rm IR}}{v}
    (v,v)
    \ .
    \end{aligned}
 \end{align}
Pulling out the factor $\omega_{\rm IR}/v$ clearly explains the nomenclature:
Apart from this overall scale the graviton scalings are exactly as in the dynamical region~\cite{Driesse:2024feo}.

As argued earlier, we want to focus only on radiative gravitons.
In this region, $\cI_{(111)}$ reduces to $\cI_{(101)}/\omega_{\rm IR}$, reflecting the fact that both types of correlator become equal in this limit.
This can be easily seen: Essentially, the radiative scaling of the zero-frequency gravitons does not require any powers of $v$.
Instead, in this region, one can just naively take the limit $v_2^\mu\to v_1^\mu$. 
Cutting the third propagator, it becomes clear that the second propagator turns into a constant factor of $\omega_{\rm IR}$.
Further the second boundary integral $\cI_{(101)}$ is a simple (massive) tadpole after cutting the third propagator.

Having determined the radiative boundary conditions, the dissipative solutions of the integrals read,
\begin{align}\label{eq:oneLoopInts}
    \vec{\mathcal{I}}
    _{\rm dis}(\gamma,\epsilon)
    =
    \frac{i}{4\pi}
    \begin{pmatrix}
        (\gamma^2-1)^{-\frac12}\, \text{arccosh}\gamma
        \\
        1
    \end{pmatrix}
    +
    \cO(\epsilon)
    \ .
\end{align}
In contrast to the PM integrals for the impulse, we see the non-trivial $\text{arccosh}\gamma$ function appearing from a one-loop diagram.

In concluding this section, we focus on the interpretation of our results in the context of BMS symmetry.
Before we indulge in this discussion, let us add a word of caution since, as was pointed out in \cite{DeAngelis:2025vlf}, it is a matter of further research to which extent BMS symmetries survive in the framework of post-Minkowskian scattering.
The angular momentum flux we are calculating can be identified with the \textit{mechanical} angular momentum from \cite{Bini:2022wrq, Riva:2023xxm}.
The loss of mechanical angular momentum is made up of two parts: The first part is due to genuine radiation of the two-body system.
The second part corresponds to the wave memory.
In our formalism, these are the dynamical and static parts respectively.

These are non-redundant effects experienced by the black holes. In particular, the mechanical angular momentum is invariant under BMS supertranslations.
Gauge redundancies arise when considering the \textit{Bondi} angular momentum, obtained from the waveform, which adds to these two contributions a term related to the shear of the gravitational field. 
BMS supertranslations alter the value of this term. Here different BMS frames, related by supertranslations, yield different results and may remove the static contribution. 
However, the Bondi angular momentum in the intrinsic frame is equal to the mechanical angular momentum while in the canonical frame it coincides with the dynamical part only \cite{Bini:2024rsy}.

By the same counting arguments as discussed in connection with eq.~\eqref{eq:countingDim}, we see that the appearance of static contributions is a peculiarity in dimensions $D\leq 4$.
In the case $l=0$ (zero incoming dynamical gravitons, i.e. Eq.~\eqref{eq:countingDim}) in general dimensions we find 
\begin{align}
m\omega_{\mathrm{IR}}^{-2}\big(
    (G m \omega_{\mathrm{IR}}^{D-3})+(G m \omega_{\mathrm{IR}}^{D-3})^2+\dots
    \big)
    \, .
\end{align}
In this expression, all terms are subleading to $\omega^{-1}$ when $D>4$.
This aligns with the findings in Ref.~\cite{Hollands:2016oma}: Static contributions are related to the wave memory, and in dimensions $D>4$ memory effects are subleading.
Interestingly, in the context of BMS symmetry, effects of supertranslations can also be shown to be negligible at leading order (in $D>4$). 
Therefore, in this case, the redundancies plaguing the Bondi angular momentum associated with them disappear at leading order. 
Thus stronger conditions on the asymptotic fall-off can be imposed, restricting the symmetry group at future null infinity to the Poincaré group $\mathbf{\mathcal{R}}^{1,D-1}\rtimes O(1,D-1)$.

\sec{Static Loss of Angular Momentum}
\label{sec:StaticAMFlux}
 
In this section we shall demonstrate our previous discussion on static contributions by computing the static impact parameter kick and loss of angular momentum up to 3PM order. We start off with a pedagogical review of 2PM and add more complexity as we arrive at 3PM.

\begin{figure}[H]
    \centering
    \begin{subfigure}{.11\textwidth}
        \centering
        \begin{tikzpicture}[baseline={(0,.4)}]
            \drawGravitonLine{0,0}{0,.9}
            \drawGravitonLine{.585,0}{.585,.9}
            \drawStaticGravitonLine{.62,0}{.62,.9}
            \drawDottedLineR{0,0}{-.4}
            \drawDottedLineR{.6,0}{.4}
            \drawDottedLineR{0,.9}{-.4}
            \drawDottedLineR{.6,0}{-.6}
            \drawStaticBHLineR{0,.91}{.6}
            \drawBHLineR{0,.89}{.6}
            \drawStaticBHLineR{.6,.91}{.4}
            \drawBHLineR{.6,.89}{.4}
            \drawPerturbativeVertex{0,0}
            \drawPerturbativeVertex{0,.9}
            \drawPerturbativeVertex{.6,0}
            \drawPerturbativeVertex{.6,.9}
        \end{tikzpicture}
        \caption{}
    \end{subfigure}
    \begin{subfigure}{.11\textwidth}
        \centering
        \begin{tikzpicture}[baseline={(0,.4)}]
            \drawGravitonLine{0,0}{.3,.45}
            \drawGravitonLine{.6,0}{.3,.45}
            \drawGravitonLine{.3,.45}{.3,.9}
            \drawDottedLineR{0,0}{-.4}
            \drawDottedLineR{.6,0}{.4}
            \drawDottedLineR{.3,.9}{-.7}
            \drawDottedLineR{.6,0}{-.6}
            \drawBHLineR{.3,.9}{.3}
            \drawBHLineR{.6,.9}{.4}
            \drawPerturbativeVertex{0,0}
            \drawPerturbativeVertex{.6,0}
            \drawPerturbativeVertex{.3,.9}
            \drawPerturbativeVertex{.3,.45}
        \end{tikzpicture}
        \caption{}
    \end{subfigure}
    \par\medskip
    \begin{subfigure}{.11\textwidth}
        \centering
        \begin{tikzpicture}[baseline={(0,.4)}]
            \drawGravitonLine{0,0}{0,.9}
            \drawGravitonLine{.585,0}{.585,.9}
            \drawStaticGravitonLine{.62,0}{.62,.9}
            \drawDottedLineR{0,0}{-.4}
            \drawDottedLineR{.6,0}{.4}
            \drawDottedLineR{.6,.9}{-1}
            \drawStaticBHLineR{0,.01}{.6}
            \drawBHLineR{0,-.01}{.6}
            \drawStaticBHLineR{.6,.91}{.4}
            \drawBHLineR{.6,.89}{.4}
            \drawPerturbativeVertex{0,0}
            \drawPerturbativeVertex{0,.9}
            \drawPerturbativeVertex{.6,0}
            \drawPerturbativeVertex{.6,.9}
        \end{tikzpicture}
        \caption{}
    \end{subfigure}
    \begin{subfigure}{.11\textwidth}
        \centering
        \begin{tikzpicture}[baseline={(0,.4)}]
            \drawGravitonLine{-.4,0}{-.4,.9}
            \drawDottedLineR{-.4,0}{-.4}
            \drawDottedLineR{-.4,0}{1.4}
            \drawDottedLineR{.6,.9}{-.6}
            \drawStaticBHLineR{.6,.9}{.4}
            \drawDottedLineR{-.4,.9}{-.4}
            \drawStaticBHLineR{0,.9}{-.4}
            \drawPerturbativeVertex{-.4,0}
            \drawPerturbativeVertex{-.4,.9}
            \drawStaticPerturbativeVertex{0,.9}
            \drawStaticPerturbativeVertex{.6,.9}
            \drawStaticCurvedGravitonLine[2]{0,.9}{.6,.9}{-80}{-100}
        \end{tikzpicture}
        \caption{}
    \end{subfigure}
    \begin{subfigure}{.11\textwidth}
        \centering
        \begin{tikzpicture}[baseline={(0,.4)}]
            \drawGravitonLine{.3,0}{.3,.5}
            \drawGravitonLine{.3,.5}{0,.9}
            \drawGravitonLine{.3,.5}{.6,.9}
            \drawDottedLineR{.3,0}{-.7}
            \drawDottedLineR{.3,0}{.7}
            \drawDottedLineR{.6,.9}{-1}
            \drawBHLineR{.6,.9}{.4}
            \drawPerturbativeVertex{0,.9}
            \drawPerturbativeVertex{.3,0}
            \drawPerturbativeVertex{.3,.5}
            \drawPerturbativeVertex{.6,.9}
        \end{tikzpicture}
        \caption{}
    \end{subfigure}
    \caption{
    The first row are ``geodesic'' diagrams scaling as $m_1^0$ while the second line is subleading in the mass and scales as $m_1^1$. 
    Gravitons and worldlines with both black and blue coloring may be either dynamic or static.
    Further, only diagrams (c) and (d) may be static and radiative.
   }
    \label{fig:2pmdiagrams}
\end{figure}
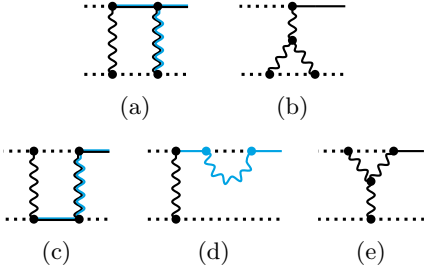

The one-point correlator $\langle z^\mu(\omega)\rangle$ is given by the sum over all diagrams with an outgoing worldline deflection.
All five diagrams contributing at $\mathcal{O}(G^2)$ are drawn in Fig.~\ref{fig:2pmdiagrams}.
Unlike the impulse we also have to take into account diagrams which have a self-interaction directly preceding the outgoing worldline (diagram (d)) as also observed in Ref.~\cite{Mogull:2025cfn} for computing the full trajectory.

At the 2PM order, it is known from the impulse that the gravitons cannot go on-shell in the dynamical region.
Therefore any dissipative effects will originate from static contributions and we have,
\begin{align}
\Delta b^{(2\mathrm{PM})\mu}_{i,\mathrm{dis}}= \Delta b^{(2\mathrm{PM})\mu}_{i,\mathrm{stat.dis.}}\, .
\end{align}
Of the five diagrams in Fig.~\ref{fig:2pmdiagrams}, only three have the structure required by the discussion in Sec.~\ref{sec:ZFandStaticCorrelators} and only two of those have dissipative contributions, namely diagrams (c) and (d).

We therefore only require the dissipative static contributions of diagrams (c) and (d) and following the discussion in section \ref{sec:ZFandStaticCorrelators}, they factorize as follows:
\begin{subequations}
 \begin{align}\label{2PMstaticEq1}
    \begin{tikzpicture}[baseline={(0,.4)}]
        \drawGravitonLine{0,0}{0,.9}
        \drawGravitonLine{.6,0}{.6,.9}
        \drawDottedLineR{0,0}{-.4}
        \drawDottedLineR{.6,0}{.4}
        \drawDottedLineR{.6,.9}{-1}
        \drawBHLineR{0,0}{.6}
        \drawBHLineR{.6,.9}{.4}
        \drawPerturbativeVertex{0,0}
        \drawPerturbativeVertex{0,.9}
        \drawPerturbativeVertex{.6,0}
        \drawPerturbativeVertex{.6,.9}
    \end{tikzpicture}
    \Bigg|_{\rm stat}
    \! \! &=
    \begin{tikzpicture}[baseline={(0,.4)}]
        \drawGravitonLine{0,0}{0,.9}
        \drawDottedLineR{0,0}{-.4}
        \drawDottedLineLR{0,.9}{0,.9}{.4}
        \drawBHLineR{0,0}{.4}
        \drawPerturbativeVertex{0,0}
        \drawPerturbativeVertex{0,.9}
        \node[below] at (.4,0) {$\nu$};
    \end{tikzpicture}
    \ 
    \int_{\omega^\prime}    
    \begin{tikzpicture}[baseline={(0,.4)}]
         \coordinate (ca) at (.2,0);
    \coordinate (cb) at (1,.9);
        \drawStaticGravitonLine{.6,0}{.6,.9}
        \drawDottedLineR{.6,0}{.4}
        \drawDottedLineR{.6,.9}{-.4}
        \drawStaticBHLineR{.6,.9}{.4}
        \drawStaticBHLineR{.6,0}{-.4}
         \node[left] at (ca) {$z^\nu(\omega')$};
        \node[right] at (cb) {$z^\mu(\omega_{\rm IR})$};
        \drawStaticPerturbativeVertex{.6,0}
        \drawStaticPerturbativeVertex{.6,.9}
    \end{tikzpicture}
    \ 
\\
 \begin{tikzpicture}[baseline={(0,.4)}]
        \drawGravitonLine{-.4,0}{-.4,.9}
        \drawDottedLineR{-.4,0}{-.4}
        \drawDottedLineR{-.4,0}{1.4}
        \drawDottedLineR{.6,.9}{-.6}
        \drawBHLineR{.6,.9}{.4}
        \drawDottedLineR{-.4,.9}{-.4}
        \drawBHLineR{0,.9}{-.4}
        \drawPerturbativeVertex{-.4,0}
        \drawPerturbativeVertex{-.4,.9}
         \drawPerturbativeVertex{0,.9}
        \drawPerturbativeVertex{.6,.9}
        \drawCurvedGravitonLine{0,.9}{.6,.9}{-80}{-80}
    \end{tikzpicture}
    \Bigg|_{\rm stat}
    \! \! &=
    \begin{tikzpicture}[baseline={(0,.4)}]
        \drawGravitonLine{0,0}{0,.9}
        \drawDottedLineR{0,0}{-.4}
        \drawBHLine{0,.9}{.4,.9}
        \drawDottedLineR{0,0}{.4}
        \drawDottedLineR{0,.9}{-.4}
        \drawPerturbativeVertex{0,0}
        \drawPerturbativeVertex{0,.9}
        \node[below] at (.4,.9) {$\nu$} ;
    \end{tikzpicture}
    \int_{\omega^\prime}    
    \!
    \begin{tikzpicture}[baseline={(0,.4)}]
    \coordinate (ca) at (-.4,.45);
    \coordinate (cb) at (1.2,.45);
        \drawDottedLine{.8,.45}{0,.45}
        \drawStaticBHLine{0,.45}{-.4,.45}
        \drawStaticBHLine{.8,.45}{1.2,.45}
        \node[left] at (ca) {$z^\nu(\omega^\prime)$};
         \node[right] at (cb) {$z^\mu(\omega_{\rm IR})$};
        \drawStaticPerturbativeVertex{0,.45}
        \drawStaticPerturbativeVertex{.8,.45}
        \drawStaticCurvedGravitonLine{0,.45}{.8,.45}{-80}{-100}
    \end{tikzpicture}
\label{2PMstaticEq2}
\end{align}
\end{subequations}
Here, the dynamic factors (black) are amputated with the external energy put to zero while the static factors (blue) are not amputated with static incoming energy $\omega'$ upon which one integrates.
The dynamical factor of each line simply reduces to the 1PM impulse of black hole 2 or black hole 1 in the first or second line respectively.
The 1PM impulses are then contracted with two-point static correlators (introduced generally in Eqs.~\eqref{eq:staticCors}):
\begin{subequations}
\begin{align}
D_{12}^{\mu;\nu}
&=\omega_{\rm IR}^{1-2\epsilon}
\!\int_{\omega^\prime}    \!\!
    \ 
    \ 
    \begin{tikzpicture}[baseline={(0,.4)}]
        \coordinate (ca) at (.2,0);
    \coordinate (cb) at (1,.9);
        \drawStaticGravitonLine{.6,0}{.6,.9}
        \drawDottedLineR{.6,0}{.4}
        \drawDottedLineR{.6,.9}{-.4}
        \drawStaticBHLineR{.6,.9}{.4}
        \drawStaticBHLineR{.6,0}{-.4}
        \node[left] at (ca) {$z^\nu(\omega')$};
        \node[right] at (cb) {$z^\mu(\omega_{\rm IR})$};
        \drawStaticPerturbativeVertex{.6,0}
        \drawStaticPerturbativeVertex{.6,.9}
    \end{tikzpicture}\, ,
    \ 
    \\
 D_{11}^{\mu;\nu}
 &=\omega_{\rm IR}^{1-2\epsilon}
 \!\int_{\omega^\prime}    \!\!
    \ 
    \ 
 \begin{tikzpicture}[baseline={(0,.4)}]
    \coordinate (ca) at (-.4,.45);
    \coordinate (cb) at (1.4,.45);
        \drawDottedLine{1,.45}{0,.45}
        \drawStaticBHLine{0,.45}{-.4,.45}
        \drawStaticBHLine{1,.45}{1.4,.45}
        \node[left] at (ca) {$z^\nu(\omega^\prime)$};
         \node[right] at (cb) {$z^\mu(\omega_{\rm IR})$};
        \drawStaticPerturbativeVertex{0,.45}
        \drawStaticPerturbativeVertex{1,.45}
        \drawStaticCurvedGravitonLine{0,.45}{1,.45}{-80}{-100}
    \end{tikzpicture}\, .
    \ 
\end{align}
\end{subequations}
Again, external propagators are not amputated and the integration is taken over the energy $\omega^\prime$ of the \textit{ingoing} worldline (time flowing from left to right).

Let us now put our diagrammatic equations into formulae: 
The (dissipative) static contribution of the diagram in eq.~\eqref{2PMstaticEq1} is given by $-\Delta p_{2,\lambda} D_{12}^{\mu;\lambda}$ and in the case of the second diagram~\eqref{2PMstaticEq2} we need to compute $-\Delta p_{1, \lambda} D_{11}^{\mu;\lambda}$. 
By virtue of
\begin{align}
    \label{pConservation}
\Delta p_1^\mu=-\Delta p_2^\mu
+\cO(G^3)\ ,
\end{align}
which holds true at the leading 1PM and 2PM orders, we can write the sum of the static parts of both diagrams as
\begin{align}
    \label{dpandcorrelator}
    \Delta b_{1,\mathrm{dis}}^\mu=- D^{\mu;\lambda}\, \Delta p_{1,\lambda}^{(1\mathrm{PM})}
    \ ,
\end{align}
with the combined correlator,
\begin{align}
    D^{\mu;\nu}=D_{11}^{\mu;\nu}-D_{12}^{\mu;\nu}\ .
\end{align}
Note that the minus sign in Eq.~\eqref{dpandcorrelator} comes about from a combination of the overall factors of Eqs.~\eqref{eq:impulse} and Eq.~\eqref{eq:DeltaB} for the relation of the impulse and impact parameter kick to the WQFT one-point function.

Inserting Feynman rules and using the integral results of Eq.~\eqref{2PMstaticEq1} we arrive at an explicit result for $D^{\mu;\nu}$.
In particular, its orthogonal components are given by,
\begin{align}\label{eq:DOrthogonal}
    &D^{\mu;\nu}_\perp=
    P^{\mu}_{\perp\alpha}
    P^{\nu}_{\perp\beta}
    D^{\alpha;\beta}
    =
    -\frac{G}{2}\mathcal{I}(\gamma)P_{\perp}^{\mn}\, ,
\end{align}
where we encounter $\cI(\gamma)$ given by,
\begin{align}
    &\mathcal{I}(\gamma)
    =
    \frac23\frac{8-5\gamma^2}{\gamma^2-1}
    -
    2\gamma\frac{3-2\gamma^2}{(\gamma^2-1)^{3/2}}
    \text{arccosh}\gamma\, .
\end{align}
Since the 1PM impulse is proportional to $b_\perp^\mu$, only the orthogonal components of $D^{\mu;\nu}$ are required for the 2PM calculation.
The functional dependence of Eq.~\eqref{eq:DOrthogonal} is encapsulated by the function $\mathcal{I}(\gamma)$ characteristic of the 2PM loss of angular momentum~\cite{Damour:2020tta}.

It is now straightforward to obtain the dissipative part of $\Delta b^\mu_1$ at 2PM order by inserting Eq.~\eqref{eq:DOrthogonal} into \eqref{dpandcorrelator}. 
The change in the orthogonal impact parameter may then easily be assembled and we find:
\begin{align}\label{eq:static2PM}
\Delta b^{(2\mathrm{PM})\mu}_{\perp,\mathrm{dis}}
=
-\frac{ G^2 m_1 m_2 2(2\gamma^2-1)\mathcal{I}(\gamma)\hat b^\mu
}{
    b\sqrt{\gamma^2-1}}\, ,
\end{align}
From this result one may in turn directly compute the 2PM loss of angular momentum,
\begin{align}
    \Delta L
    =
    -\frac{G^2 M^2\mu^2}{E b}
    2(2\gamma^2-1)
        \mathcal{I}(\gamma)
        \ ,
\end{align}
in agreement with the result first derived by Damour~\cite{Damour:2020tta}.

As a step up from 2PM, we turn to the static contribution to $\Delta b_{i,\mathrm{dis}}^\mu$ at $\cO(G^3)$.
\begin{figure*}[t]
\begin{center}
    \begin{subfigure}{.13\textwidth}
        \centering
        \begin{tikzpicture}[baseline={(0,.4)}]
            \drawGravitonLine{0,0}{0,.9}
            \drawGravitonLine{.6,0}{.6,.9}
            \drawStaticGravitonLine{1.2,0}{1.2,.9}
            \drawDottedLineR{0,0}{-.4}
            \drawStaticBHLineR{.6,0}{.6}
            \drawBHLine{.6,.9}{0,.9}
            \drawDottedLineR{1.2,0}{.4}
            \drawDottedLineR{0,0}{.6}
            \drawDottedLineR{.6,.9}{.6}
            \drawDottedLine{0,.9}{-.4,.9}
            \drawStaticBHLineR{1.2,.9}{.4}
            \drawPerturbativeVertex{0,0}
            \drawPerturbativeVertex{0,.9}
            \drawPerturbativeVertex{.6,0}
            \drawPerturbativeVertex{.6,.9}
            \drawStaticPerturbativeVertex{1.2,.9}
            \drawStaticPerturbativeVertex{1.2,0}
        \end{tikzpicture}
        \caption{}
    \end{subfigure}
    \begin{subfigure}{.14\textwidth}
        \centering
        \begin{tikzpicture}[baseline={(0,.4)}]
            \drawDottedLineR{-1,0}{-.4}
            \drawDottedLineR{-1,.9}{-.4}
            \drawDottedLineR{-1,0}{.6}
            \drawDottedLine{-.4,.0}{1,0}
            \drawDottedLine{0,.9}{.6,.9}
            \drawStaticBHLineR{.6,.9}{.4}
            \drawBHLine{-1,.9}{-.4,.9}
            \drawGravitonLine{-1,.9}{-1,0}
            \drawGravitonLine{-.4,.9}{-.4,0}
            \drawStaticBHLineR{0,.9}{-.4}
            \drawPerturbativeVertex{-.4,0}
            \drawStaticPerturbativeVertex{0,.9}
            \drawStaticPerturbativeVertex{.6,.9}
            \drawPerturbativeVertex{-.4,.9}
            \drawPerturbativeVertex{-1,0}
            \drawPerturbativeVertex{-1,.9}
            \drawStaticCurvedGravitonLine[2]{0,.9}{.6,.9}{-80}{-100}
        \end{tikzpicture}
        \caption{}
    \end{subfigure}
        \begin{subfigure}{.13\textwidth}
        \centering
        \begin{tikzpicture}[baseline={(0,.4)}]
            \drawGravitonLine{.3,.4}{.3,.9}
            \drawGravitonLine{.3,.4}{0,.0}
            \drawGravitonLine{.3,.4}{.6,.0}
            \drawStaticGravitonLine{1.2,0}{1.2,.9}
            \drawDottedLine{.3,.9}{1.2,.9}
            \drawDottedLine{0,0}{.6,0}
            \drawDottedLine{.3,.9}{-.4,.9}
            \drawStaticBHLine{.6,.0}{1.2,0}
            \drawDottedLine{0,0}{-.4,0}
            \drawDottedLineR{1.2,0}{.4}
            \drawStaticBHLineR{1.2,.9}{.4}
            \drawPerturbativeVertex{.3,.4}
            \drawPerturbativeVertex{0,0}
            \drawPerturbativeVertex{.3,.9}
            \drawPerturbativeVertex{.6,.0}
            \drawStaticPerturbativeVertex{1.2,.9}
            \drawStaticPerturbativeVertex{1.2,0}
        \end{tikzpicture}
        \caption{}
    \end{subfigure}
    \begin{subfigure}{.13\textwidth}
        \centering
        \begin{tikzpicture}[baseline={(0,.4)}]
            \drawDottedLineR{-.7,.9}{-.7}
            \drawDottedLineR{-1,0}{-.4}
            \drawDottedLineR{-1,0}{.6}
            \drawDottedLine{-.3,.0}{.7,0}
            \drawDottedLine{-.3,.9}{.2,.9}
            \drawStaticBHLineR{.3,.9}{.4}
            \drawGravitonLine{-.7,.9}{-.7,.4}
            \drawGravitonLine{-.7,.4}{-1,0}
            \drawGravitonLine{-.7,.4}{-.4,0}
            \drawStaticBHLineR{-.3,.9}{-.4}
            \drawPerturbativeVertex{-.4,0}
            \drawStaticPerturbativeVertex{-.3,.9}
            \drawStaticPerturbativeVertex{.3,.9}
            \drawPerturbativeVertex{-.7,.9}
            \drawPerturbativeVertex{-1,0}
            \drawPerturbativeVertex{-.7,.4}
            \drawStaticCurvedGravitonLine[2]{-.3,.9}{.3,.9}{-80}{-100}
        \end{tikzpicture}
        \caption{}
    \end{subfigure}
    \begin{subfigure}{.14\textwidth}
        \centering
        \begin{tikzpicture}[baseline={(0,.4)}]
            \drawGravitonLine{-.4,0}{-.4,.9}
            \drawGravitonLine{-1,0}{-1,.9}
            \drawStaticCurvedBHLine{-1,.9}{0,.9}{40}{140}
            \drawDottedLineR{-.4,0}{1.4}
            \drawDottedLineR{.6,.9}{-.6}
            \drawDottedLine{-1,.9}{-1.4,.9}
            \drawDottedLine{-1,0}{-1.4,0}
            \drawStaticBHLineR{.6,.9}{.4}
            \drawDottedLine{-1,0}{-.4,0}
            \drawStaticBHLineR{0,.9}{-.4}
            \drawPerturbativeVertex{-.4,0}
            \drawPerturbativeVertex{-.4,.9}
            \drawStaticPerturbativeVertex{0,.9}
            \drawStaticPerturbativeVertex{.6,.9}
            \drawPerturbativeVertex{-1,0}
            \drawPerturbativeVertex{-1,.9}
            \drawStaticCurvedGravitonLine[2]{0,.9}{.6,.9}{-80}{-100}
        \end{tikzpicture}
        \caption{}
    \end{subfigure}
    \begin{subfigure}{.13\textwidth}
        \centering
        \begin{tikzpicture}[baseline={(0,.4)}]
            \drawGravitonLine{0,0}{0,.9}
            \drawGravitonLine{.6,0}{.6,.9}
            \drawStaticGravitonLine{1.2,0}{1.2,.9}
            \drawDottedLineR{0,0}{-.4}
            \drawStaticBHLineR{.6,0}{.6}
            \drawStaticCurvedBHLine{0,.9}{1.2,.9}{40}{140}
            \drawDottedLineR{1.2,0}{.4}
            \drawDottedLineR{0,0}{.6}
            \drawDottedLine{0,.9}{-.4,.9}
            \drawStaticBHLineR{1.2,.9}{.4}
            \drawPerturbativeVertex{0,0}
            \drawPerturbativeVertex{0,.9}
            \drawPerturbativeVertex{.6,0}
            \drawPerturbativeVertex{.6,.9}
            \drawStaticPerturbativeVertex{1.2,.9}
            \drawStaticPerturbativeVertex{1.2,0}
            \drawDottedLineR{.6,.9}{.6}
        \end{tikzpicture}
        \caption{}
    \end{subfigure}
    \begin{subfigure}{.13\textwidth}
        \centering
        \begin{tikzpicture}[baseline={(0,.4)}]
            \drawGravitonLine{0,0}{0,.9}
            \drawGravitonLine{.9,0}{.9,.9}
            \drawDottedLineR{0,0}{-.4}
            \drawStaticCurvedGravitonLine{.3,.9}{1.2,.9}{-80}{-100}
            \drawDottedLine{.3,.9}{.9,.9}
            \drawDottedLine{.9,0}{1.6,0}
            \drawDottedLineR{0,0}{.9}
            \drawDottedLine{0,.9}{-.4,.9}
            \drawStaticBHLine{.9,.9}{1.2,.9}
            \drawStaticBHLine{0,.9}{.3,.9}
            \drawStaticBHLineR{1.2,.9}{.4}
            \drawStaticPerturbativeVertex{.3,.9}
            \drawPerturbativeVertex{0,0}
            \drawPerturbativeVertex{0,.9}
            \drawPerturbativeVertex{.9,0}
            \drawPerturbativeVertex{.9,.9}
            \drawStaticPerturbativeVertex{1.2,.9}
        \end{tikzpicture}
        \caption{}
    \end{subfigure}
\end{center}
\begin{center}
    \begin{subfigure}{.14\textwidth}
        \centering
        \begin{tikzpicture}[baseline={(0,.4)}]
            \drawGravitonLine{0,0}{0,.9}
            \drawGravitonLine{.6,0}{.6,.9}
            \drawStaticGravitonLine{1.2,0}{1.2,.9}
            \drawDottedLineR{0,0}{-.4}
            \drawDottedLineR{.6,.9}{.6}
            \drawDottedLineR{.6,.9}{-1}
            \drawDottedLineR{1.2,0}{.4}
            \drawBHLineR{0,0}{.6}
            \drawStaticBHLineR{.6,0}{.6}
            \drawStaticBHLineR{1.2,.9}{.4}
            \drawPerturbativeVertex{0,0}
            \drawPerturbativeVertex{0,.9}
            \drawPerturbativeVertex{.6,0}
            \drawPerturbativeVertex{.6,.9}
            \drawStaticPerturbativeVertex{1.2,.9}
            \drawStaticPerturbativeVertex{1.2,0}
        \end{tikzpicture}
        \caption{}
    \end{subfigure}
    \begin{subfigure}{.14\textwidth}
        \centering
        \begin{tikzpicture}[baseline={(0,.4)}]
            \drawGravitonLine{-.4,0}{-.4,.9}
            \drawGravitonLine{-1,0}{-1,.9}
            \drawDottedLineR{-.4,.9}{-.6}
            \drawDottedLineR{-.4,0}{1.4}
            \drawDottedLineR{.6,.9}{-.6}
            \drawDottedLineR{-.4,.9}{-.4}
            \drawDottedLine{-1,.9}{-1.4,.9}
            \drawDottedLine{-1,0}{-1.4,0}
            \drawStaticBHLineR{.6,.9}{.4}
            \drawBHLine{-1,0}{-.4,0}
            \drawStaticBHLineR{0,.9}{-.4}
            \drawPerturbativeVertex{-.4,0}
            \drawPerturbativeVertex{-.4,.9}
            \drawStaticPerturbativeVertex{0,.9}
            \drawStaticPerturbativeVertex{.6,.9}
            \drawPerturbativeVertex{-1,0}
            \drawPerturbativeVertex{-1,.9}
            \drawStaticCurvedGravitonLine[2]{0,.9}{.6,.9}{-80}{-100}
        \end{tikzpicture}
        \caption{}
    \end{subfigure}
    \begin{subfigure}{.14\textwidth}
        \centering
        \begin{tikzpicture}[baseline={(0,.4)}]
            \drawGravitonLine{.3,0}{.3,.5}
            \drawGravitonLine{.3,.5}{0,.9}
            \drawGravitonLine{.3,.5}{.6,.9}
            \drawStaticGravitonLine{1.2,0}{1.2,.9}
            \drawDottedLineR{.3,0}{-.7}
            \drawDottedLineR{.6,.9}{.6}
            \drawDottedLineR{.6,.9}{-1}
            \drawDottedLineR{1.2,0}{.4}
            \drawStaticBHLineR{.3,0}{.9}
            \drawStaticBHLineR{1.2,.9}{.4}
            \drawPerturbativeVertex{.3,.5}
            \drawPerturbativeVertex{0,.9}
            \drawPerturbativeVertex{.3,0}
            \drawPerturbativeVertex{.6,.9}
            \drawStaticPerturbativeVertex{1.2,.9}
            \drawStaticPerturbativeVertex{1.2,0}
        \end{tikzpicture}
        \caption{}
    \end{subfigure}
    \begin{subfigure}{.14\textwidth}
        \centering
        \begin{tikzpicture}[baseline={(0,.4)}]
            \drawDottedLineR{-.4,.9}{-.6}
            \drawDottedLineR{-.7,0}{1.7}
            \drawDottedLineR{.6,.9}{-.6}
            \drawDottedLineR{-.4,.9}{-.4}
            \drawDottedLine{-1,.9}{-1.4,.9}
            \drawDottedLine{-.7,0}{-1.4,0}
            \drawStaticBHLineR{.6,.9}{.4}
            \drawGravitonLine{-.7,0}{-.7,.5}
            \drawGravitonLine{-.7,.5}{-1,.9}
            \drawGravitonLine{-.7,.5}{-.4,.9}
            \drawStaticBHLineR{0,.9}{-.4}
            \drawPerturbativeVertex{-.4,.9}
            \drawStaticPerturbativeVertex{0,.9}
            \drawStaticPerturbativeVertex{.6,.9}
            \drawPerturbativeVertex{-.7,0}
            \drawPerturbativeVertex{-1,.9}
            \drawPerturbativeVertex{-.7,.5}
            \drawStaticCurvedGravitonLine[2]{0,.9}{.6,.9}{-80}{-100}
        \end{tikzpicture}
        \caption{}
    \end{subfigure}
    \quad
    \begin{subfigure}{.14\textwidth}
        \centering
        \begin{tikzpicture}[baseline={(0,.4)}]
            \drawGravitonLine{0,0}{0,.9}
            \drawGravitonLine{.6,.18}{.6,.9}
            \drawStaticGravitonLine{1.2,0}{1.2,.9}
            \drawDottedLineR{0,0}{-.4}
            \drawStaticCurvedBHLine{.6,.18}{1.2,0}{0}{160}
            \drawStaticBHLine{0,0}{1.2,0}{-40}{-140}
            \drawDottedLineR{0,.9}{.6}
            \drawDottedLineR{.6,.9}{.6}
            \drawDottedLineR{1.2,0}{.4}
            \drawDottedLine{0,.9}{-.4,.9}
            \drawStaticBHLineR{1.2,.9}{.4}
            \drawPerturbativeVertex{0,0}
            \drawPerturbativeVertex{0,.9}
            \drawPerturbativeVertex{.6,.18}
            \drawPerturbativeVertex{.6,.9}
            \drawStaticPerturbativeVertex{1.2,.9}
            \drawStaticPerturbativeVertex{1.2,0}
        \end{tikzpicture}
        \caption{}
    \end{subfigure}
\end{center}
\caption{
    Diagrams constituting the static loss of angular momentum at 3PM order. 
    The first row of diagrams (a)-(g) contribute at 1SF order and the second row (h)-(l) at 2SF order.
    Static gravitons and worldline deflections are drawn in blue.
    All diagrams with two 1PM insertions (g) and (j)-(l) drop out in the total loss of angular momentum.
}
\end{figure*}
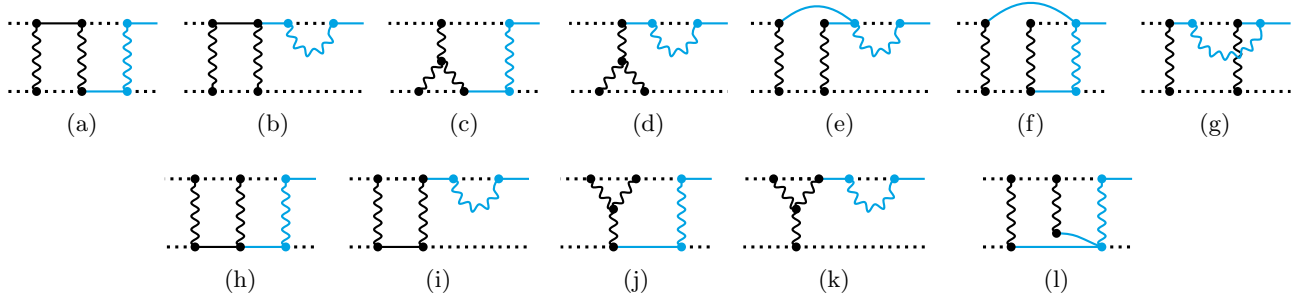
At this order, the dissipative impact parameter kick is no longer solely contained in the static region. 
Instead, the dynamical region contributes as well:
\begin{align}
    \Delta b^{(3\mathrm{PM})\mu}_{1,\mathrm{dis}}
    =
    \Delta b^{(3\mathrm{PM})\mu}_{1,\mathrm{stat.dis.}}
    +
    \Delta b^{(3\mathrm{PM})\mu}_{1,\mathrm{rad}}\, .
\end{align}
The dissipative dynamical part is simply labelled as ``radiative'' (because it corresponds to ``real'' radiation of gravitational waves).
We postpone its discussion to the next section and focus first on the static dissipative contribution.
We now pose the question of which static correlators of Eq.~\eqref{generalstaticSnakesMushrooms} play a role at 3PM order and identify two subsets, distinguished by the number of impulse insertions in the diagram.

Firstly, we can combine a static two-point correlator with one insertion of the 2PM impulse. For this calculation we may recycle our result of $D^{\mu;\nu}$ from the previous section and compute
\begin{align}
    -D^{\mu;\lambda}
    \Delta p_{1,\lambda}^{(2\mathrm{PM})} 
    \ .
\end{align}
At this point, however, we require the velocity components of $D^{\mu\lambda}$ in addition to the orthogonal components given in Eq.~\eqref{eq:DOrthogonal}.

Secondly, one can combine static three-point correlators with two insertions of the 1PM impulse.
In particular, there are four three-point static correlators:
\begin{align}\label{eq:ThreeCorrelators}
    \begin{tikzpicture}[baseline={(0,.4)}]
        \drawStaticGravitonLine{.6,0}{.6,.9}
        \drawDottedLineR{.6,0}{.4}
        \drawStaticBHLineR{.6,.9}{-.4}
        \drawStaticBHLineR{.6,.9}{.4}
        \drawStaticBHLineR{.6,0}{-.4}
        \drawStaticPerturbativeVertex{.6,0}
        \drawStaticPerturbativeVertex{.6,.9}
    \end{tikzpicture}
    \ ,
    \quad
     \begin{tikzpicture}[baseline={(0,.4)}]
        \drawStaticGravitonLine{.6,0}{.6,.9}
        \drawDottedLineR{.6,0}{.4}
        \drawDottedLineR{.6,.9}{-.4}
        \drawStaticBHLineR{.6,.9}{.4}
        \drawStaticBHLineR{.6,0}{-.4}
        \drawStaticCurvedBHLine{.6,0}{.2,.3}{160}{0}
        \drawStaticPerturbativeVertex{.6,0}
        \drawStaticPerturbativeVertex{.6,.9}
    \end{tikzpicture}
    \ ,
    \quad
    \begin{tikzpicture}[baseline={(0,.4)}]
        \drawDottedLine{0,.5}{.8,.5}
        \drawStaticBHLineR{0,.5}{-.4}
        \drawStaticBHLineR{.8,.5}{.4}
        \drawStaticCurvedBHLine{.8,.5}{.2,.8}{160}{0}
        \drawStaticPerturbativeVertex{0,.5}
        \drawStaticPerturbativeVertex{.8,.5}
        \drawStaticCurvedGravitonLine{0,.5}{.8,.5}{-80}{-100}
    \end{tikzpicture}
    \ ,
    \quad
    \begin{tikzpicture}[baseline={(0,.4)}]
        \drawDottedLine{0,.5}{.8,.5}
        \drawStaticBHLineR{0,.5}{-.4}
        \drawStaticBHLineR{.8,.5}{.4}
        \drawStaticCurvedBHLine{0,.5}{-.4,.8}{160}{0}
        \drawStaticPerturbativeVertex{0,.5}
        \drawStaticPerturbativeVertex{.8,.5}
        \drawStaticCurvedGravitonLine{0,.5}{.8,.5}{-80}{-100}
    \end{tikzpicture}
    \ .
\end{align}
After contraction with the two factors of the 1PM impulse, the contribution from the three-point correlators is proportional only to the velocities.
Thus, for the physical, projected change in the impact parameter, these terms drop out and only the first contributions with a single 2PM insertion are relevant.
This, we anticipate, will  change at higher orders in $G$.

One may directly argue that the 3PM contributions from the three-point correlators must be proportional to $v_i^\mu$.
Namely, the correlators themselves are three-index tensors composed only of $v_i^\mu$ and $\eta_\mn$.
Two of the three indices are contracted with factors of $b_\perp^\mu$ from the 1PM impulse.
The only possibility is $b_\perp^\mu b_\perp^\nu \eta_\mn v_i^\sigma$ because $b_\perp$ is orthogonal to $v_i^\mu$.

The final result of the total (physically projected) static, dissipative impact parameter kick at 3PM order takes the form,
\begin{align}\label{eq:static3PM}
&\Delta b_{\perp,{\rm stat.dis.}}^{(3\mathrm{PM})\mu}=
-\frac{G^3 M^2\mu \mathcal{I}(\gamma )}{4 b^2} \Big[\frac{3 \pi  (5 \gamma 
^2-1) \hat{b}^\mu}{\sqrt{\gamma ^2-1}}\\
&+\frac{16 (1-2 
\gamma ^2)^2 ((m_1+\gamma  m_2) v_2^\mu- 
(\gamma m_1+m_2)v_1^\mu)}{M(\gamma ^2-1)^2}\Big]\, .
\nn
\end{align}
The 3PM static dissipation of angular momentum resulting from this result is presented in Sec.~\ref{sec:ResGR}.

Notably, the projected kick at this order gets two contributions which schematically may be written as,
\begin{align}
    \sim{G\,\cI(\gamma)
    \Big(
        P_{\perp}^\mn \Delta p_{1\nu}^{(2\mathrm{PM})}
        +
        \Delta P_{\perp}^{(1\mathrm{PM})\mn}
        \Delta p_{1\nu}^{(1\mathrm{PM})}
    \Big)
    }
    \ .
\end{align}
The first term is the 3PM (static) impact parameter kick projected with $P_\perp^\mn$ while the second term takes the 1PM change of the projector $\Delta P_{\perp}^{(1\mathrm{PM})\mn}$ combined with the 2PM impact parameter kick into account.
The impulses appear through their insertion in the static correlators which in turn gives rise to the characteristic $\cI(\gamma)$.

\sec{Dynamic Loss of Angular Momentum}
\label{sec:DynamicLoss}
\begin{figure*}[t]
   \begin{center}
    \begin{subfigure}{.14\textwidth}
        \centering
   \begin{tikzpicture}[baseline={(0,.4)}]
        \drawGravitonLine{0,0}{0,.9}
        \drawRadiativeGravitonLine{.6,0}{.6,.9}
        \drawGravitonLine{1.2,0}{1.2,.9}
        \drawDottedLineR{0,0}{-.4}
        \drawDottedLineR{.6,0}{.6}
        \drawDottedLineR{.6,.9}{-1}
        \drawDottedLineR{1.2,0}{.4}
        \drawBHLineR{0,0}{.6}
        \drawBHLineR{.6,.9}{.6}
        \drawBHLineR{1.2,.9}{.4}
        \drawPerturbativeVertex{0,0}
        \drawPerturbativeVertex{0,.9}
        \drawPerturbativeVertex{.6,0}
        \drawPerturbativeVertex{.6,.9}
        \drawPerturbativeVertex{1.2,.9}
        \drawPerturbativeVertex{1.2,0}
    \end{tikzpicture}
    \caption{}
\end{subfigure}
\begin{subfigure}{.14\textwidth}
        \centering
    \begin{tikzpicture}[baseline={(0,.4)}]
        \drawGravitonLine{0,0}{0,.9}
        \drawRadiativeGravitonLine{.6,0}{.6,.9}
        \drawGravitonLine{1.2,0}{1.2,.9}
        \drawDottedLineR{0,0}{-.4}
        \drawBHLineR{.6,0}{.6}
        \drawBHLine{.6,.9}{0,.9}
        \drawDottedLineR{1.2,0}{.4}
        \drawDottedLineR{0,0}{.6}
        \drawDottedLineR{.6,.9}{.6}
        \drawDottedLine{0,.9}{-.4,.9}
        \drawBHLineR{1.2,.9}{.4}
        \drawPerturbativeVertex{0,0}
        \drawPerturbativeVertex{0,.9}
        \drawPerturbativeVertex{.6,0}
        \drawPerturbativeVertex{.6,.9}
        \drawPerturbativeVertex{1.2,.9}
        \drawPerturbativeVertex{1.2,0}
    \end{tikzpicture}
    \caption{}
\end{subfigure}
\begin{subfigure}{.14\textwidth}
        \centering
    \begin{tikzpicture}[baseline={(0,.4)}]
        \drawGravitonLine{0,0}{0,.9}
        \drawGravitonLine{.6,0}{.6,.9}
        \drawRadiativeGravitonLine{1.2,0}{1.2,.9}
        \drawDottedLineR{0,0}{-.4}
        \drawBHLineR{.6,0}{.6}
        \drawCurvedBHLine{0,.9}{1.2,.9}{40}{140}
        \drawDottedLineR{1.2,0}{.4}
        \drawDottedLineR{0,0}{.6}
        \drawDottedLine{0,.9}{-.4,.9}
        \drawBHLineR{1.2,.9}{.4}
        \drawPerturbativeVertex{0,0}
        \drawPerturbativeVertex{0,.9}
        \drawPerturbativeVertex{.6,0}
        \drawPerturbativeVertex{.6,.9}
        \drawPerturbativeVertex{1.2,.9}
        \drawPerturbativeVertex{1.2,0}
        \drawDottedLineR{.6,.9}{.6}
    \end{tikzpicture}
    \caption{}
\end{subfigure}
\begin{subfigure}{.14\textwidth}
        \centering
    \begin{tikzpicture}[baseline={(0,.4)}]
        \drawGravitonLine{0,0}{0,.9}
        \drawGravitonLine{.9,0}{.9,.9}
        \drawDottedLineR{0,0}{-.4}
        \drawRadiativeCurvedGravitonLine{.3,.9}{1.2,.9}{-80}{-100}
        \drawDottedLine{.3,.9}{.9,.9}
        \drawDottedLine{.9,0}{1.6,0}
        \drawDottedLineR{0,0}{.9}
        \drawDottedLine{0,.9}{-.4,.9}
        \drawBHLine{.9,.9}{1.2,.9}
        \drawBHLine{0,.9}{.3,.9}
        \drawBHLineR{1.2,.9}{.4}
        \drawPerturbativeVertex{.3,.9}
        \drawPerturbativeVertex{0,0}
        \drawPerturbativeVertex{0,.9}
        \drawPerturbativeVertex{.9,0}
        \drawPerturbativeVertex{.9,.9}
        \drawPerturbativeVertex{1.2,.9}
    \end{tikzpicture}
    \caption{}
\end{subfigure}
\begin{subfigure}{.14\textwidth}
        \centering
   \begin{tikzpicture}[baseline={(0,.4)}]
        \drawGravitonLine{0,0}{0,.9}
        \drawGravitonLine{1.2,0}{1.2,.9}
        \drawDottedLineR{0,0}{-.4}
        \drawDottedLineR{.3,0}{.9}
        \drawDottedLineR{0,.9}{-.4}
        \drawDottedLineR{1.2,0}{.4}
        \drawDottedLineR{0,0}{.3}
        \drawDottedLineR{0,.9}{1.2}
        \drawBHLineR{1.2,.9}{.4}
        \drawBHLineR{0,0}{.3}
        \drawBHLineR{.9,0}{.3}
        \drawRadiativeCurvedGravitonLine{.3,0}{.9,0}{80}{100}
        \drawPerturbativeVertex{0,0}
        \drawPerturbativeVertex{0,.9}
        \drawPerturbativeVertex{.3,0}
        \drawPerturbativeVertex{1.2,.9}
        \drawPerturbativeVertex{.9,0}
        \drawPerturbativeVertex{1.2,0}
    \end{tikzpicture}
    \caption{}
\end{subfigure}
\begin{subfigure}{.14\textwidth}
        \centering
    \begin{tikzpicture}[baseline={(0,.4)}]
        \drawGravitonLine{0,0}{0,.9}
        \drawGravitonLine{1.2,0}{1.2,.9}
        \drawDottedLineR{0,0}{-.4}
        \drawDottedLineR{.3,0}{.9}
        \drawDottedLineR{0,.9}{-.4}
        \drawDottedLineR{1.2,0}{.4}
        \drawDottedLineR{0,0}{.3}
        \drawDottedLineR{0,.9}{1.2}
        \drawBHLineR{1.2,.9}{.4}
        \drawBHLineR{0,.9}{.3}
        \drawBHLineR{.9,.9}{.3}
        \drawRadiativeCurvedGravitonLine{.3,.9}{.9,.9}{-80}{-100}
        \drawPerturbativeVertex{0,0}
        \drawPerturbativeVertex{0,.9}
        \drawPerturbativeVertex{.3,.9}
        \drawPerturbativeVertex{1.2,.9}
        \drawPerturbativeVertex{.9,.9}
        \drawPerturbativeVertex{1.2,0}
\end{tikzpicture}
\caption{}
\end{subfigure}
\begin{subfigure}{.14\textwidth}
        \centering
    \begin{tikzpicture}[baseline={(0,.4)}]
        \drawGravitonLine{0,0}{0,.9}
        \drawGravitonLine{.9,0}{.9,.5}
        \drawRadiativeGravitonLine{.9,.5}{.6,.9}
        \drawGravitonLine{.9,.5}{1.2,.9}
        \drawDottedLine{.6,.9}{1.2,.9}
        \drawDottedLineR{0,0}{-.4}
        \drawDottedLineR{0,.9}{-.4}
        \drawDottedLineR{.9,0}{.7}
        \drawDottedLineR{0,0}{.9}
        \drawBHLineR{0,.9}{.6}
        \drawBHLineR{1.2,.9}{.4}
        \drawPerturbativeVertex{0,0}
        \drawPerturbativeVertex{0,.9}
        \drawPerturbativeVertex{.9,0}
        \drawPerturbativeVertex{.9,.5}
        \drawPerturbativeVertex{.6,.9}
        \drawPerturbativeVertex{1.2,.9}
    \end{tikzpicture}
    \caption{}
\end{subfigure}
\begin{subfigure}{.14\textwidth}
        \centering
    \begin{tikzpicture}[baseline={(0,.4)}]
        \drawGravitonLine{0,0}{0,.9}
        \drawGravitonLine{.9,0}{.9,.5}
        \drawRadiativeGravitonLine{.9,.5}{1.2,.9}
        \drawGravitonLine{.9,.5}{.6,.9}
        \drawDottedLineR{0,0}{-.4}
        \drawCurvedBHLine{0,.9}{1.2,.9}{40}{140}
        \drawDottedLineR{.9,0}{.7}
        \drawDottedLineR{0,0}{.9}
        \drawDottedLine{0,.9}{-.4,.9}
        \drawBHLineR{1.2,.9}{.4}
        \drawPerturbativeVertex{0,0}
        \drawPerturbativeVertex{0,.9}
        \drawPerturbativeVertex{.9,0}
        \drawPerturbativeVertex{.9,.5}
        \drawPerturbativeVertex{.6,.9}
        \drawPerturbativeVertex{1.2,.9}
        \drawDottedLineR{.6,.9}{.6}
    \end{tikzpicture}
    \caption{}
\end{subfigure}
\begin{subfigure}{.14\textwidth}
        \centering
    \begin{tikzpicture}[baseline={(0,.4)}]
        \drawGravitonLine{.3,0}{.3,.5}
        \drawGravitonLine{.3,.5}{0,.9}
        \drawRadiativeGravitonLine{.3,.5}{.6,.9}
        \drawGravitonLine{1.2,0}{1.2,.9}
        \drawDottedLine{0,.9}{-.4,.9}
        \drawDottedLine{0,.9}{.6,.9}
        \drawDottedLine{.3,0}{-.4,0}
        \drawDottedLine{.3,.0}{1.2,0}
        \drawDottedLineR{1.2,0}{.4}
        \drawBHLineR{1.2,.9}{.4}
        \drawBHLine{.6,.9}{1.2,.9}
        \drawPerturbativeVertex{.3,.5}
        \drawPerturbativeVertex{0,.9}
        \drawPerturbativeVertex{.3,0}
        \drawPerturbativeVertex{.6,.9}
        \drawPerturbativeVertex{1.2,.9}
        \drawPerturbativeVertex{1.2,0}
    \end{tikzpicture}
    \caption{}
\end{subfigure}
\begin{subfigure}{.14\textwidth}
        \centering
    \begin{tikzpicture}[baseline={(0,.4)}]
        \drawGravitonLine{0,0}{0,.9}
        \drawRadiativeGravitonLine{.6,0}{.9,.4}
        \drawGravitonLine{.9,.4}{.9,.9}
        \drawGravitonLine{.9,.4}{1.2,0}
        \drawDottedLineR{0,0}{-.4}
        \drawDottedLineR{0,.9}{-.4}
        \drawDottedLine{0,.9}{.9,.9}
        \drawDottedLineR{.6,0}{1.2,0}
        \drawBHLineR{0,0}{.6}
        \drawBHLineR{.9,.9}{.7}
        \drawPerturbativeVertex{0,0}
        \drawPerturbativeVertex{0,.9}
        \drawPerturbativeVertex{.9,.9}
        \drawPerturbativeVertex{.9,.4}
        \drawPerturbativeVertex{.6,0}
        \drawPerturbativeVertex{1.2,0}
    \end{tikzpicture}
    \caption{}
\end{subfigure}
\begin{subfigure}{.14\textwidth}
        \centering
    \begin{tikzpicture}[baseline={(0,.4)}]
        \drawGravitonLine{.3,.4}{.3,.9}
        \drawGravitonLine{.3,.4}{0,.0}
        \drawRadiativeGravitonLine{.3,.4}{.6,.0}
        \drawGravitonLine{1.2,0}{1.2,.9}
        \drawDottedLine{.3,.9}{1.2,.9}
        \drawDottedLine{0,0}{.6,0}
        \drawDottedLine{.3,.9}{-.4,.9}
        \drawBHLine{.6,.0}{1.2,0}
        \drawDottedLine{0,0}{-.4,0}
        \drawDottedLineR{1.2,0}{.4}
        \drawBHLineR{1.2,.9}{.4}
        \drawPerturbativeVertex{.3,.4}
        \drawPerturbativeVertex{0,0}
        \drawPerturbativeVertex{.3,.9}
        \drawPerturbativeVertex{.6,.0}
        \drawPerturbativeVertex{1.2,.9}
        \drawPerturbativeVertex{1.2,0}
    \end{tikzpicture}
    \caption{}
\end{subfigure}
\begin{subfigure}{.14\textwidth}
        \centering
    \begin{tikzpicture}[baseline={(0,.4)}]
        \drawGravitonLine{0,0}{0,.9}
        \drawRadiativeGravitonLine{0,.45}{1.2,.45}
        \drawGravitonLine{1.2,0}{1.2,.9}
        \drawDottedLineR{0,0}{-.4}
        \drawDottedLineR{0,.9}{1.2}
        \drawDottedLineR{0,.9}{-.4}
        \drawDottedLineR{1.2,0}{.4}
       \drawDottedLineR{0,0}{1.2}
        \drawBHLineR{1.2,.9}{.4}
        \drawPerturbativeVertex{0,0}
        \drawPerturbativeVertex{0,.9}
        \drawPerturbativeVertex{0,.45}
        \drawPerturbativeVertex{1.2,.45}
        \drawPerturbativeVertex{1.2,.9}
        \drawPerturbativeVertex{1.2,0}
    \end{tikzpicture}
    \caption{}
\end{subfigure}
    \end{center}
    \caption{
        Diagrams contributing to the dynamical loss of angular momentum at 3PM order. 
        Only 1SF diagrams are present in the dynamical region.
        Each diagram has exactly one radiative graviton indicated by red color.
        }
        \label{fig:3pmgraphs}
\end{figure*}
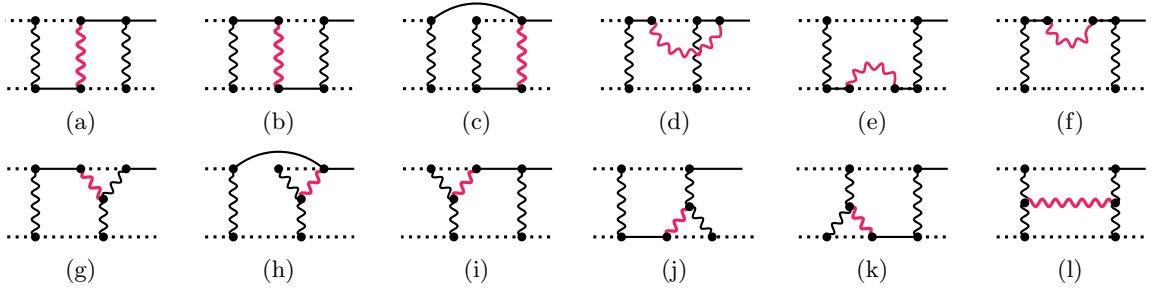
We now turn to the dynamical region. 
The integrand is obtained by assuming the dynamical scaling $\ell^\mu\sim (q,q)$ for all graviton momenta and $\omega\sim q$ for all WL frequencies. 
As described in Sec.~\ref{sec:angularMomentum}, one may use Eq.~\eqref{eq:DeltaB} which we reprint here for clarity:
\begin{align}
    \Delta b^\mu_{i,{\rm dis}}
    =
    \lim_{\omega_{\rm IR}\to0}
    \frac{\d}{\d\omega_{\rm IR}}
    i\omega_{\rm IR}^2
    \langle 
        z_i^\mu(\omega_{\rm IR})
    \rangle
    \bigg|_{\rm dis}
    \ .
\end{align}
With the dynamical scalings, the application of this formula mirrors exactly a computation of the impulse where the only difference would be omitting the derivative on $\omega_{\rm IR}$.
In the dynamical region, the only effect of this derivative is to alter slightly the Feynman rules associated with the outgoing line.
In particular, the relevant WQFT diagrams are the same except we need only the subset describing dissipative effects.
The relevant diagrams are shown in Fig.~\ref{fig:3pmgraphs}.

Since the only difference to an impulse computation lies in the outgoing vertex rule, the propagator structure remains unaltered and we encounter exactly the same kind of integrals.
These are (classical) two-loop integrals whose evaluation is well established~\cite{Bern:2019nnu,Herrmann:2021lqe,Jakobsen:2022fcj,Jakobsen:2022psy}.
They may generally be parametrized by an integral family with seven indices labelling the powers of seven propagator structures,
\begin{align}
    I_{(\nu_1\dots\nu_7)}(\gamma,\epsilon,q)
    =
    \int_{\ell_1, \ell_2} 
    \frac{
        2\pi\delta(\ell_1\cdot v_{1})2\pi\delta(\ell_2\cdot v_2)
        }{
            \prod_{i=1}^{7}D_{i}^{\nu_i}
        }\, ,
\end{align}
with the propagator structures $D_i$ given by,
\begin{align}
    D_1=\ell_1\cdot v_2
    \,,\ 
    D_2=\ell_2\cdot v_1
    \,,\
    D_3=(k^0+i0)^2-\vct{k}^2
    \,,\ 
    \nn
    \\
    D_4=\ell_1^2
    \,,\ 
    D_5=\ell_2^2
    \,,\ 
    D_6=(\ell_1-q)^2
    \,,\ 
    D_7=(\ell_2-q)^2
    \ ,
    \label{eq:Propgs}
\end{align}
and with $k^\mu=\ell_1^\mu-\ell_2^\mu$.
In particular, the third propagator involving $k^\mu$ is the ``active graviton'' highlighted in red in the diagrams of Fig.~\ref{fig:3pmgraphs}.
The factors of $2\pi$ in front of the delta functions are for convenience.

Since we are interested only in dissipative effects, we require only contributions to the two-loop integrals where the active graviton is radiative.
This way of identifying radiative effects at the 3PM order is well-established~\cite{Herrmann:2021lqe,Jakobsen:2022psy}.
Among others, this restriction to radiative effects allows us to neglect the $i0$ propagator pole displacements of the linear propagators as they will never play a role~\cite{Jakobsen:2023hig}.
We may then label the integrals such that the active graviton $D_3$ always comes out with the causal retarded prescription given in Eq.~\eqref{eq:Propgs}.

From computations of the impulse it is useful to categorize the integrals into $v$-type integrals with odd $(\nu_1+\nu_2)$ accompanying $v_i^\mu$ and $b$-type integrals with even $(\nu_1+\nu_2)$ accompanying $b^\mu$ (or $q^\mu$ in Fourier space) --- see e.g. Ref.~\cite{Driesse:2024feo,Jakobsen:2022zsx} though the exact terminology varies.
For the impact parameter kick, the association is reversed: The $v_i^\mu$ direction comes with $b$-type integrals and the $b^\mu$ direction with $v$-type integrals.
To a certain degree, the $b$-type integrals mainly describe conservative effects and the $v$-type integrals mainly describe dissipative effects and indeed the $b^\mu$ direction of the impact parameter kick (associated with $v$-type integrals) describes the main effects of dissipation of $\Delta b_i^\mu$.

Indeed, the 3PM dynamical loss of angular momentum is only sensitive to the $b^\mu$ direction of the impact parameter kick and for this result we therefore only require dissipative $v$-type integrals.
After tensor reduction and IBP reduction, we find four scalar master integrals given by,
\begin{align}
    \vec{I}_{\rm MIs}(\gamma,\epsilon)
    =
    q^{4\epsilon}
    \begin{pmatrix}
    \ q^{\hphantom{3}} I_{(-1,0,1,1,0,1,0)}(\gamma,\epsilon,q)
    \\[3pt]
    \ q^{\hphantom{3}} I_{(0,-1,1,1,0,0,1)}(\gamma,\epsilon,q)
    \\[3pt]
    \ q^{3} I_{(+1,0,1,1,0,0,1)}(\gamma,\epsilon,q)
    \\[3pt]
    \ q^{5} I_{(0,-1,1,1,1,1,1)}(\gamma,\epsilon,q)
    \end{pmatrix}
    \ .
\end{align}
In Ref.~\cite{Jakobsen:2022psy}, they are computed with retarded boundary conditions using the method of differential equations.
There only total expressions including both potential and radiative gravitons are given.
It is, however, straightforward to separate out the radiative part and including only dissipative effects (defined by fixing boundary conditions in the PN limit in the radiative region), we find:
\begin{align}
    \vec{I}_{\rm MIs,dis}(\gamma,\epsilon)
    =
    \frac{i}{32\pi}
    \begin{pmatrix}
        -\tfrac18 (\gamma^2-1)
        \\[5.5pt]  
        -\tfrac12 \frac{\gamma-1}{\gamma+1}
        \\[5.5pt]
        (\gamma^2-1)^{-1/2}\text{arccosh}\gamma
        \\[5.5pt]
        \text{log}\frac{\gamma+1}{2}
    \end{pmatrix}
    +
    \cO(\epsilon)
    \ .
\end{align}
Interestingly, all integrals are finite and we require their expressions only to leading order in $\epsilon$.
Note also that the fourth master integral is a ``non-linear'' effect (i.e. requires self-interactions in the bulk) and does not appear in EM.
In effect, as we will see, the EM results do not include the logarithm associated with that master integral.

The function space of the master integrals $\vec{\cI}_{\rm MIs,dis}$ is encapsulated in the following vector:
\begin{align}\label{eq:fncSpace}
    \vec{I}(\gamma)
    =
    \begin{pmatrix}
        1
        \\[5pt]
        (\gamma^2-1)^{-1/2}\text{arccosh}\gamma
        \\[5pt]
        \text{log}\frac{\gamma+1}2
    \end{pmatrix}
    \ .
\end{align}
We may therefore expand all our dynamic, dissipative two-loop results on this basis of functions.

Carrying out the computation in analogy to the 3PM calculation of the impulse~\cite{Jakobsen:2022psy,Jakobsen:2022fcj}, we arrive at our result for the dynamical dissipative contributions to the two-loop impact parameter kick:
\begin{align}\label{eq:impactParameterKick}
&
\Delta b_{\perp{\rm dyn.dis.}}^{(3\mathrm{PM})\mu}
=
\frac{G^3 M^2\mu}{48 b^2} \bigg[\frac{\pi  \hat b^\mu}{(\gamma 
^2\!-\!1)^{5/2}} 
\vec{f}(\gamma)\cdot\vec{I}(\gamma)
\\
&-\frac{96 (1-2 \gamma ^2)^2 
\big[(\gamma m_1+m_2)v_1^\mu-(m_1+\gamma  m_2) v_2^\mu\big]}
    {
    M(\gamma^2-1)^2
    }
    \mathcal{I}(\gamma ) 
    \bigg]\, .
\nn
\end{align}
Focusing on the orthogonal impact parameter $b_{\perp}^\mu$ requires only the $v$-type integrals discussed above.
The $v$-direction of this result comes from the dissipative two-loop change in the projector $P_\perp^\mn$.
The $b$ direction has been expanded on the function space $\vec{I}$ with expansion coefficients $\vec{f}(\gamma)=(f_1(\gamma),f_2(\gamma),f_3(\gamma))$ given by,

\begin{align}
\nn &f_1(\gamma)=270 \gamma ^6-141 \gamma ^5-1986 \gamma ^4+4514 \gamma^3
\\ \nn
&\qquad\qquad
-3558 \gamma ^2+379 \gamma +474\  ,
\\[3pt] 
& f_2(\gamma)=3 \gamma  (2\gamma ^2-3)
\\ \nn
&
\times\big(105 \gamma ^5-120 \gamma ^4-170 \gamma ^3
+144\gamma ^2+49 \gamma -24\big)\, ,
\\[3pt] \nn
&f_3(\gamma)=-6 (\gamma ^2-1) 
\\ \nn
&
\times\big(105 \gamma ^5-120 \gamma^4-290 \gamma ^3+108 \gamma ^2
+305 \gamma -124\big)\, .
\end{align}
The orthogonal impact parameter kick determines the loss of angular momentum if the total loss of four momentum is known.
However, inverting that relationship we may also determine the 3PM change in the orthogonal impact parameter and doing so, we have checked explicitly our result~\eqref{eq:impactParameterKick}.

A more direct check of the velocity direction of the impact parameter kick is also offered by its property of being orthogonal to the velocities and reads,
\begin{align}\label{eq:check1}
    0
    &=
    (b_\perp + \Delta b_\perp)
    \cdot
    (p_{i}+\Delta p_{i})
    \\
    &=
    \Delta b_\perp\cdot p_i
    +
    b_\perp\cdot\Delta p_i
    +
    \Delta b_\perp\cdot \Delta p_i
    \ .
    \nn
\end{align}
In other words, the orthogonal impact parameter at future infinity is orthogonal to the impulses at future infinity.
In fact, since the conservative contributions to $\Delta b^\mu_\perp$ and $\Delta p^\mu_i$ satisfy Eq.~\eqref{eq:check1} by themselves, one may discard them and focus only on dissipative contributions.
Further, we note that static contributions also obey Eq.~\eqref{eq:check1} by themselves --- at least to the 3PM order relevant here.
With this restriction only the two first terms of the second line of Eq.~\eqref{eq:check1} contribute for the dynamical region.
This check then explains why the velocity direction of $\Delta b_{\perp\rm dyn.dis.}^{(3PM)\mu}$ must be proportional to $\cI(\gamma)$: This function appears at 3PM order via linear response in the radiative terms of the $b$-direction $\Delta p_i$ which therefore appears in the constraint~\eqref{eq:check1}.
We have also checked that the static contributions Eqs.~\eqref{eq:static2PM} and~\eqref{eq:static3PM} satisfy the check~\eqref{eq:check1}.

\sec{Results in Gravity}
\label{sec:ResGR}
In this section we present our result for the
change in angular momentum in gravity.
In addition we present also the total change in the four-momentum of the two black holes which we computed as a step towards the angular momentum loss.


As discussed around Eq.~\eqref{eq:EL} we parametrize the loss of angular momentum through the change in $EL$ which we expand as,
\begin{align}\label{deltaEL}
    &\Delta (E L)
    =
    \frac{G^2 M^2\mu^2}{b}
    \Big[
        c_{\rm stat}^{(2)}(\gamma)
        \mathcal{I}(\gamma)
        +
        \frac{GM}{b} 
        c_{\rm stat}^{(3)}(\gamma)
        \mathcal{I}(\gamma)
        \nn\\
    &\qquad\qquad\qquad
        +
        \frac{GM}{b}
        \vec{c}^{\ (3)}_{\rm dyn}(\gamma)\cdot \vec{I}(\gamma)
    \Big]
        +
        \cO(G^4)
    \ ,
\end{align}
with the static part parametrized by the expansion coefficients $c_{\rm stat}^{(n)}(\gamma)$ and the dynamic part by $\vec{c}_{\rm dyn}^{\ (3)}(\gamma)$ with the function space vector $\vec{I}(\gamma)$ from Eq.~\eqref{eq:fncSpace}.
The coefficients are provided explicitly below in Eqs.~\eqref{eq:coefficients}.
Continuing, we expand the change in the total energy as,
\begin{align}\label{deltaE}
    &\Delta E
    =
    \!\frac{G^3 M^3\mu^2}{b^3 E}\!
    \vec{d}_{\rm dyn}^{\ (3)}(\gamma)\cdot \vec{I}(\gamma)+
    \cO(G^4)
    \ ,
\end{align}
which only has dynamic contributions parametrized by $\vec{d}_{\rm dyn}^{\ (3)}(\gamma)$ which is also provided in Eqs.~\eqref{eq:coefficients}.

The change in angular momentum $\Delta L$ is a simple linear combination of the above two results:
\begin{align}\label{eq:lalala}
    \Delta L
    =
    \frac1{E}
    \Delta(EL)
    -
    \frac{L}{E}
    \Delta E
    +
    \cO(G^5)
    \ .
\end{align}
The reason for splitting it into these terms is the mass dependence: Each term has its own distinct mass dependence and one would generally not expect simplifications between them.
More specifically the difference in mass dependence is hidden in the variables $L$ from Eq.~\eqref{eq:l} and $E^2=m_1^2+m_2^2+2\gamma m_1 m_2$.
At $\cO(G^5)$ another contribution to Eq.~\eqref{eq:lalala} appears which combines $\Delta E$ and $\Delta L$.

Finally, we may parametrize the vectorial total fluxes of linear and angular momentum in terms of the scalar losses:
\begin{align}
    L_{\rm dis}^\mu
    =
    -\Delta L^\mu 
    &= 
    -\Delta L
    \,
    \hat L^\mu
    \ ,
    \\
    P_{\rm rad}^\mu
    =-
    \Delta P^\mu
    &=
    -\frac{
        E \Delta E
    }{M}
    \frac{v_1^\mu+v_2^\mu}{1+\gamma}
    +
    \cO(G^4)
    \ .
\end{align}
Here, we provided also the radiative loss carried away by the gravitational field which is simply the negative of the change of each of the vectors.
For the angular momentum loss we used as label ``dissipative'' instead of ``radiative'' because of the ambiguity of whether to interpret the static loss of angular momentum as radiation.
At higher PM orders, it is no longer possible simply to parametrize the total loss of four-momentum in terms only of the energy loss as there are also independent recoil effects.

The static expansion coefficients of the angular momentum loss are relatively simple and read:
\begin{subequations}\label{eq:coefficients}
\begin{align}
&c_{\rm stat}^{(2)}=-2(2\gamma^2-1)\, ,
\\
&c_{\rm stat}^{(3)}=-\frac{3\pi}{4}(5\gamma^2-1)\, .
\end{align}
Instead, the two-loop dynamical vector coefficients are more involved in their structure and read,
\begin{widetext}
\begin{align}
    &\vec{c}_{\rm dyn}^{\ (3)}
    =
    \frac{\pi}{24(\gamma^2-1)^2}
    \begin{pmatrix}
        -105\gamma^7+411\gamma^6-240\gamma^5-537\gamma^4+683\gamma^3-111\gamma^2-386\gamma+237
        \\[5pt]
        3\gamma(2\gamma^2-3)(35 \gamma^5\!-\!60\gamma^4\!-\!70\gamma^3\!+\!72\gamma^2\!+\!19\gamma\!-\!12)
        \\[5pt]
        -6(\gamma^2-1)(35\gamma^5\!-\!90\gamma^4\!-\!70\gamma^3\!+\!16\gamma^2\!+\!155\gamma\!-\!62)
    \end{pmatrix}\ ,
    \\[6pt]
    &\vec{d}_{\rm dyn}^{\ (3)}
    =
    \frac{\pi}{48(\gamma^2-1)^{5/2}}
    \begin{pmatrix}
        (\gamma^2-1)(-210\gamma^6+552\gamma^5-339\gamma^4+912\gamma^3-3148\gamma^2+3336\gamma-1151)
        \\[5pt]
        -3(\gamma^2-1)(70\gamma^6-165\gamma^4+112\gamma^2-33)
        \\[5pt]
        210\gamma^8+360\gamma^7-1320\gamma^6-264\gamma^5+1980\gamma^4-552\gamma^3-840\gamma^2+456\gamma-30
    \end{pmatrix}\ .
\end{align}
\end{widetext}
\end{subequations}
Apart from the overall factors, the entries of both coefficient vectors are polynomials in $\gamma$.

As we have already stated, our results in GR are already well known~\cite{Damour:2020tta,Jakobsen:2021smu,Manohar:2022dea,DiVecchia:2022piu,Herrmann:2021lqe,Herrmann:2021tct}.
In particular, our 2PM static result for $\Delta L$ reproduces Ref.~\cite{Damour:2020tta} and the 3PM part (static and dynamic) of $\Delta L$ reproduces Ref.~\cite{Manohar:2022dea}.
Finally, our 3PM result for $\Delta E$ reproduces Ref.~\cite{Herrmann:2021lqe}.

\sec{Results in Electromagnetism}
\label{sec:ResultEM}
The action for the gauge potential in electromagnetism is,
\begin{align}
    S_{\mathrm{EM}}
    =
    -
    \tfrac14 
    \int \d^D x F_{\mu\nu}(x)F^{\mu\nu}(x)\, .
\end{align}
In the WEFT framework, we couple it to point-like particles which are now parametrized by their mass $m_i$ and charge $e q_i$ and their worldline action is,
\begin{align}
    S_{m_i}
    =
    -
    \int \d\tau 
    \left[
        \tfrac{1}{2}
        m_i
        \dot{x}_i^\mu
        \dot{x}_i^\nu
        \eta_{\mn}
        +
        e q_i A_\mu(x_i)
        \dot x_i^\mu
    \right]\, .
\end{align}
Here, we dropped the constant term in the Brink - Di Vecchia - Howe part of the action and $-e=-|e|$ is the electron charge.

We introduce the fine structure constant,
\begin{align}
    \alpha=\frac{e^2}{4\pi}\ ,
\end{align}
which in this context plays a parallel role to $G$ in GR.
The analogous expansion parameters to the Schwarzschild radii $G m_i$ of the BHs are the ``classical radii'' of the charged particles:
\begin{align}
    \lambda_i=\frac{(e q_i)^2}{4\pi m_i}
    =
    \alpha \frac{q_i^2}{m_i}
    \ .
\end{align}
The effective field theory description is valid as long as $\lambda_i$ are much smaller than the relevant interaction scale, in our case $b$.
The total action reads,
\begin{align}
    S=
    S_{m_1}
    +
    S_{m_2}
    +
    S_{\mathrm{EM}}
    +
    S_{\mathrm{gf}}
    \ ,
\end{align}
where we also added a gauge fixing term.
We choose Feynman gauge which in this classical context corresponds to Lorenz gauge.
The background expansion of the trajectories are carried out as in GR given by Eq.~\eqref{eq:WLExp} and the photon field $A_\mu(x)$ is considered a perturbation itself.

The change in $L$ is defined from the background parameters exactly as discussed in Sec.~\ref{sec:angularMomentum}.
Further, the WQFT methodology for computing $\Delta L$ is theory agnostic and may be carried directly over to EM.
This includes both Eq.~\eqref{eq:DeltaB} which relates $\Delta b_i^\mu$ to the WQFT one-point function and the analysis of static regions in Sec.~\ref{sec:ZFandStaticCorrelators}.

The main difference between the GR and EM computations lies only in the Feynman rules.
These are generally simpler because of $A_\mu(x)$ having only one index instead of the two indices of $h_\mn(x)$.
More importantly, however, the photon field does not interact with itself and we therefore do not have self-interactions in the bulk in EM.
Therefore, the required set of diagrams is much smaller and includes only diagrams without bulk interactions.
In the dynamic region, this restriction removes half of the diagrams and leaves only the first five diagrams of Fig.~\ref{fig:3pmgraphs}.
The number of required static correlators is the same (as we are not yet sensitive to ``non-linear memory'' effects), though the dynamic impulse insertions are simpler.

The steps for computing the EM loss of angular momentum are therefore exactly analogous to the GR computation, though only simpler.
Eventually we find the following result for $\Delta L$ parametrized through $\Delta(EL)$ given explicitly by,
\begin{widetext}
\begin{align}
&
\Delta (EL)=
\frac{\pi \alpha^2 q_1^2 q_2^2}{3 b}\bigg[\frac{4 \gamma m_2 q_1}{\pi m_1 q_2}+\frac{4 \gamma m_1 q_2}{\pi m_2 q_1}
-\frac{12\gamma \cF(\gamma)}{\pi 
(\gamma^2-1)}
-\frac{\alpha}{b}\bigg(\frac{M q_1^2}{m_1^2}+\frac{M q_2^2}{m_2^2}
-\frac{3 q_1 q_2\cF(\gamma)}{\mu 
(\gamma^2-1)}
\bigg)\bigg]
\\[4pt]
&
-\!
\frac{\pi \alpha^3 q_1^2q_2^2}{b^2(\gamma^2\!-\!1)}
\bigg[\!\left(
    \frac{m_1 q_2^2}{m_2^2} 
    +
    \frac{m_2 q_1^2}{m_1^2}
    \right)
    \frac{3\gamma^4+6\gamma^2-1}{12}
+
\left(
    \frac{q_1^2}{m_1}+\frac{q_2^2}{m_2}
    \right)\frac{2\gamma}{3}
+
\frac{q_1 q_2}{\mu} \bigg(2(\gamma-1)-\frac{(\gamma^2+2\gamma-1)\mathcal{F}(\gamma)}{(\gamma^2-1)}\bigg)
\bigg]
\!+
\mathcal{O}(\alpha^4)
\,  ,
\nn
\end{align}
\end{widetext}
where we have introduced the function,
\begin{align}
    \cF(\gamma)
    =
    \gamma-\frac{\text{arccosh}\gamma}{(\gamma^2-1)^{1/2}}\ .
\end{align}
The first line of $\Delta (EL)$ provides the $\cO(\alpha^2)$ and $\cO(\alpha^3)$ static losses while the second line gives the $\cO(\alpha^3)$ dynamical contribution.
The $\cO(\alpha^2)$ part of this result reproduces Ref.~\cite{Saketh:2021sri}.
As far as we are aware, however, the $\cO(\alpha^3)$ has not yet been covered in the literature.
Nonetheless, the full dissipative three-loop impulse in EM is known\footnote{This result is given in the master thesis by Alain Goldberg~\cite{Goldberg} provided to us via private communication with Gregor K\"alin.} and we have checked both the $\cO(\alpha^2)$ and $\cO(\alpha^3)$ parts of our result via linear response using the approach of Refs.~\cite{Jakobsen:2023hig,Jakobsen:2023pvx}.

We also reproduce the radiated energy at $\mathcal{O}(\alpha^3)$
\begin{align}\label{EMdeltaE}
    &\Delta E=\\
    &-\frac{\pi \alpha^3 q_1^2 q_2^2}{12 b^3E}\bigg[\frac{3\gamma^2\!+\!1}{\sqrt{\gamma^2\!-\!1}}\bigg(\!\frac{(m_1\!+\!\gamma m_2)q_1^2}{m_1^2}+\frac{(m_2\!+\!\gamma m_1)q_2^2}{m_2^2}\!\bigg)\nn\\
    &+\frac{3\, q_1 q_2}{\mu \sqrt{\gamma^2\!-\!1}^{3}}(4\,(\gamma^2\!-\!2\gamma\!+\!1)-(3\gamma^2\!+\!1)\cF(\gamma))\bigg]+
    \cO(\alpha^4)\nn
    \, ,
\end{align}
in agreement with \cite{Saketh:2021sri,Bern:2021xze}. 

As is evident from our calculation, radiation of angular momentum in EM scattering includes static contributions.
A natural question is whether this result is invariant under asymptotic symmetry transformations or, more generally, if it is possible to define an invariant notion of radiated angular momentum. Recall that one could define the mechanical angular momentum which is invariant under BMS supertranslations in gravity.
It was pointed out in \cite{Fuentealba:2023rvf,Cristofoli:2022phh} that the situation in electromagnetism closely mirrors that in gravity: An asymptotic u(1) gauge symmetry in EM plays a role similar to the BMS supertranslations.
As in gravity, one can define a notion of angular momentum invariant under these asymptotic transformations. 
We conjecture that the quantity we have computed can be identified with this analog of the mechanical angular momentum in electromagnetism.

\sec{Conclusions}
\label{sec:conclusion}
In this work, we propose a new approach to calculating the total angular momentum dissipated in a gravitational two-body scattering event.
We use the WQFT formalism and identify a way to extract the total change in each of the BH impact parameters and from these changes, one may then compute the total loss of angular momentum.
Most simply, this extraction is realized via a generalization of the WQFT formula for computing the BH impulses from the on-shell limit of tree-level one-point correlation functions.
Namely, the change in the BH impact parameters are derived from the subleading term in the soft expansion of the correlation functions in the outgoing frequency.

Passing to subleading order in this expansion introduces subtle effects.
First and foremost, the external frequency tending to zero plays the role of an infrared regulator and gives rise to new (from the perspective of WQFT) static integration regions which we identify with well-known zero-frequency gravitons.
We classify how this new integration region appears in generic WQFT diagrams which boils down to one-loop $n$-point static correlators.
The contributions from this region give rise to the static parts of the loss of angular momentum and, in particular, that it appears already at 2PM order.

The subtle effects of the change in the impact parameters may also be identified with an ill-definedness of its time components.
This leads to a divergent behavior of conservative contributions to these components.
However, for the computation of the loss of angular momentum we are only interested in dissipative effects.
Focusing exclusively on such effects, all divergent behavior of the change of the impact parameter disappears --- at least to the $\cO(G^3)$ order considered in this work.
While the dissipative effects are of most interest, it would be desirable to also understand the conservative effects in future work.

Intriguingly, apart from the static integration regions (well under control) and focusing only on dissipative effects, the computation of the change of the impact parameter --- and hence the loss of angular momentum --- reduces to exactly the same diagrams and integrals as for the impulse.
For the $\cO(G^3)$ calculation considered explicitly in this work we could therefore directly import the well-known two-loop integral results required at this order.

Having established this method, the next natural step is to use the powerful methods developed within WQFT to compute new results for the loss of angular momentum at high perturbative orders matching those of the impulse.
A first clear target is to complete the spinless 4PM sector where the required three-loop integrals are also well studied.
Other avenues will be to add spin and other finite-size effects.

One interesting first application of the 4PM loss of angular momentum will be to connect it, via linear response, to the linear-in-radiation part of the 5PM impulse~\cite{Driesse:2024feo}.
Further, considerable effort has gone into resumming perturbative PM results in order to reproduce numerical relativity simulations as accurately as possible~\cite{Damour:2014afa,Damour:2022ybd,Swain:2024ngs,Clark:2025kvu,Long:2025nmj}, as well as comparisons of PM results with results derived via the gravitational self-force approach~\cite{Warburton:2025ymy,Barack:2026izc}.
Here high-precision PM fluxes of angular momentum may well improve resummation models and will be an interesting target for self-force comparisons.

The 4PM loss of angular momentum will be sensitive to the non-linear memory effect.
In contrast, up to the 3PM order considered here, only the linear memory contributes.
Generally, it would be interesting to connect the static correlators and the zero-frequency-graviton analysis carried out in this paper to the small-frequency expansion of the waveform and linear and non-linear memory effects.
Concluding, the 4PM loss of angular momentum will be an exciting --- and realistic --- next step.

\medskip

\sec{Acknowledgments}
We thank Stefano De Angelis, Carlo Heissenberg, Gregor K\"alin, Jan Plefka, Jerome Reuschenbach and Niels Warburton for useful discussions.
This work was funded by the Deutsche Forschungsgemeinschaft (DFG, German Research Foundation) Projektnummer 417533893/GRK2575 “Rethinking Quantum Field Theory” (KS).

\bibliographystyle{JHEP}
\bibliography{zfRefs}

\end{document}